\documentclass[ALICE,manyauthors]{cernphprep}

\usepackage[comma,square,numbers,sort&compress]{natbib}
\usepackage{multirow}
\usepackage{siunitx}
\usepackage{comment}
\DeclareSIUnit[per-mode = fraction]\eVperc{eV/\text{\ensuremath{c}}}
\DeclareSIUnit\invbarn{b\ensuremath{^{-1}}}
\usepackage{hyperref}
\usepackage[mathlines]{lineno}
\usepackage[T1]{fontenc}
\usepackage{orcidlink}

\usepackage{xspace}
\usepackage{color}
\usepackage{xcolor}
\usepackage{rotating}

\newcommand{\sNN}{\ensuremath{\sqrt{s_{\rm NN}}}\xspace}
\newcommand{\pt}{\ensuremath{p_{\rm T}}} 
\newcommand{\dedx}{\ensuremath{{\rm d}E/{\rm d}x}\xspace}

\begin{document}
%\linenumbers
%%%%%%%%%%%%%%%  Title page %%%%%%%%%%%%%%%%%%%%%%%%
\begin{titlepage}
% the dates below correspond to CERN approval
% please don't touch: EB chairs will take care
\PHyear{2024}       % required, will be obtained from CERN
\PHnumber{142}      % required, will be obtained from CERN
\PHdate{22 May}  % required, will be obtained from CERN
%%%%%%%%%%%%%%%%%%%%%%%%%%%%%%%%%%%%%%%%%%%%%%%%%%%%

%%% Put your own title + short title here:
\title{Investigating strangeness enhancement in jet and medium via $\phi$(1020) production in p--Pb collisions at \ensuremath{\sqrt{s_{\rm NN}}} = 5.02 TeV}
\ShortTitle{Jet and medium production of $\phi$(1020) in p--Pb collisions}

%%% Do not change the next lines
\Collaboration{ALICE Collaboration\thanks{See Appendix~\ref{app:collab} for the list of collaboration members}}
\ShortAuthor{ALICE Collaboration}

\begin{abstract}
 This work aims to differentiate strangeness produced from hard processes (jet-like) and softer processes (underlying event) by measuring the angular correlation between a high-momentum trigger hadron (h) acting as a jet-proxy and a produced strange hadron ($\phi(1020)$ meson). Measuring h--$\phi$ correlations at midrapidity in p--Pb collisions at \ensuremath{\sqrt{s_{\rm NN}}} = 5.02 TeV as a function of event multiplicity provides insight into the microscopic origin of strangeness enhancement in small collision systems. The jet-like and the underlying-event-like strangeness production are investigated as a function of event multiplicity. They are also compared between a lower and higher momentum region. The evolutions of the per-trigger yields within the near-side (aligned with the trigger hadron) and away-side (in the opposite direction of the trigger hadron) jets are studied separately, allowing for the characterization of two distinct jet-like production regimes. Furthermore, the h--$\phi$ correlations within the underlying event give access to a production regime dominated by soft production processes, which can be compared directly to the in-jet production. Comparisons between h--$\phi$ and dihadron correlations show that the observed strangeness enhancement is largely driven by the underlying event, where the $\phi/\mathrm{h}$ ratio is significantly larger than within the jet regions. As multiplicity increases, the fraction of the total $\phi(1020)$ yield coming from jets decreases compared to the underlying event production, leading to high-multiplicity events being dominated by the increased strangeness production from the underlying event.
    
\end{abstract}
\end{titlepage}
\setcounter{page}{2}
  
\section{Introduction}
\label{sec:intro}
The enhancement of strange-quark production in heavy-ion collisions with respect to minimum bias proton--proton (pp) collisions has long been one of the suggested signatures of the existence of a deconfined, partonic phase of nuclear matter known as the quark--gluon plasma (QGP)~\cite{rafelski_strange}. The QGP phase is expected to be formed in extremely high-energy-density environments, such as those produced in heavy-ion collisions~\cite{qgp_heavyions}. The enhancement of strange and multi-strange hadrons in the QGP with respect to their production within a hadronic-gas phase was one of the major predicted signatures of QGP formation in heavy-ion collisions~\cite{Koch:1986ud}. The measurement of strange-quark enhancement via the production of strange hadrons has been investigated for many years in different collision systems and at different energies by experiments at the Super Proton Synchrotron (SPS), the Relativistic Heavy Ion Collider (RHIC), and the Large Hadron Collider (LHC)~\cite{strangeoverview, Khachatryan_2017, Abelev_2014}. The production of strange particles in heavy-ion collisions is successfully described by a grand canonical statistical model of particle production in an equilibrated QGP medium~\cite{heavyion_production, grandcanonical}. 

In addition to enhancement in heavy-ion collisions, more recent studies at LHC energies have measured strangeness enhancement in high-multiplicity pp and p--Pb collisions with respect to minimum bias pp collisions~\cite{ALICE:2016fzo, 2016389}. Strikingly, this enhancement, quantified by the ratio of strange to non-strange hadron (i.e., pion) yields, is seen to increase smoothly as a function of the event multiplicity, independent of the collision system. Since the enhancement in larger systems is attributed to a thermalized QGP state, this experimental evidence of strangeness enhancement in small collision systems raises questions about the early characteristics and evolution of pp and p--Pb collision systems. While certain theoretical models --- such as statistical thermal models~\cite{Cleymans_2004,Hamieh_2000}, color reconnection and rope hadronization~\cite{Bierlich_2015, Nayak_2019}, and the dynamical core-corona model~\cite{PhysRevC.101.024912, PhysRevC.79.064907} --- can qualitatively explain this behavior in small collision systems, the exact production mechanism responsible for this increase in strangeness in these systems remains unknown.

Experimentally, strangeness enhancement can be measured by looking at the ratio of strange particles to non-strange particles, most commonly pions. The increase of this yield ratio with increasing multiplicity scales with the strange-quark content of the particle species under study. Even though the $\phi$(1020) meson ($s\overline{s}$) has hidden strangeness, the ALICE Collaboration have measured an enhancement in the $\phi$(1020) production in between the increase seen for hadrons containing one and two strange quarks, such as the $\Lambda$ and the $\Xi$, respectively~\cite{PhysRevC.99.024906,ALICE:2016fzo,2016389}. In this paper, the enhancement in strange-quark production in small collision systems is investigated by studying the $\phi$(1020) production in p--Pb collisions at center-of-mass energy per nucleon pair $\mbox{\sNN{} = \SI{5.02}{\tera\eV}}$ collected during the LHC data-taking period Run 2. The event multiplicity classes in p--Pb collisions provide a valuable span of collision activity, overlapping the event multiplicity of pp collisions as well as peripheral Pb--Pb collisions. The p--Pb collision system also allows for a balance between hard scattering and soft production processes, allowing for both in-jet and underlying-event production to be investigated across a range of collision multiplicities. Therefore, the p--Pb collision system is ideal for studying the origin of strangeness enhancement.

In order to explore the mechanisms responsible for the enhancement of $\phi$(1020) meson production as a function of system size, the analysis presented in this paper separates the $\phi$(1020) production into event regions that are dominated by hard-scattering processes (i.e., jets) from those dominated by softer interactions (i.e., the underlying event). Separation of the yields corresponding to these different production regimes can be achieved through measurements of two-particle angular correlations~\cite{idcorr,PhysRevC.71.051902}. Previous studies have used both jet reconstruction and two-particle angular correlations to study strange baryon and meson production ($\Lambda^0$, K$^0_\mathrm{S}$) in jets, showing a significant difference in behavior between jet and non-jet production~\cite{starstrangejets,alicelambdajets,alicestrangejetcorr}. By looking at the angular correlation between a high-\pt{} trigger hadron, interpreted as a proxy for the jet axis, and a $\phi$(1020) meson, the $\phi$(1020) meson yield can be divided into three regions: the yield within the near-side jet (aligned with the trigger hadron), within the away-side jet (opposite the trigger hadron), and within the underlying event. Similar correlation measurements are obtained using the same trigger requirements correlated with inclusive charged hadrons (h--h) for direct comparison.

The per-trigger yield ratio of (h--$\phi$)/(h--h) pairs gives the $\phi/\mathrm{h}$ ratio within events that contain trigger hadrons. Thus, using angular correlations to extract pairwise h--$\phi$ and h--h yields in different regions of correlation space the approximate $\phi/\mathrm{h}$ production ratio is extracted within the near-side jet, the away-side jet, and the underlying-event-dominated regions. This allows one to study the multiplicity (and system size) dependence of $\phi$(1020) production in a region dominated by hard-scattering processes by measuring the $\phi/\mathrm{h}$ ratio in the jet-dominated regions, and for soft processes by measuring it in the region where the underlying event is dominant. The separated near-side and away-side jet yields are also used to investigate possible modification of the $\phi/\mathrm{h}$ ratio caused by differences in jet fragmentation and potential jet-medium interactions.

The next section, Section~\ref{sec:experiment}, describes the ALICE sub-detectors that are used for this analysis. Section~\ref{sec:data} explains the data analysis details, while Section~\ref{sec:corr} describes the specific measurement procedure in more detail, followed by a discussion of the systematic effects and corresponding uncertainties in Section~\ref{sec:syst}. Finally, the h--$\phi$ correlation results are presented in Section~\ref{sec:results}, concluding with a discussion on how these findings address the origin of strangeness enhancement.

\section{Experimental setup}
\label{sec:experiment}
The ALICE detector at the LHC consists of a central barrel encompassing several sub-detectors housed within a large solenoid magnet, which provides a magnetic field of \SI{0.5}{\tesla} in the direction of the beam axis. In addition, several forward and backward detectors are used for event characterization and triggering. A detailed description of the ALICE detector can be found in~\cite{ALICE:2008ngc,2014perf}. The primary detectors used in this analysis are the Inner Tracking System (ITS), the Time Projection Chamber (TPC), and the Time Of Flight (TOF) detector. The ITS allows an accurate reconstruction of the primary collision vertex and distinguishes between primary tracks originating from the primary vertex and secondary tracks resulting from weak decays. The TPC is a cylindrical drift chamber that acts as the main tracking detector of the ALICE apparatus, covering the central pseudorapidity region within $|\eta| < 0.9$. In addition, the TPC track information is also used for particle identification (PID) by measuring the specific energy loss \dedx{} of the tracks traversing the chamber, parameterized with the Bethe-Bloch formula. Situated outside of the TPC and covering the same $|\eta| < 0.9$ range is the TOF detector, an array of Multigap Resistive Plate Chambers (MRPC). The TOF gives precise track timing information with a resolution of \SI{56}{ps} that is used for additional PID constraints~\cite{TOFperf_run2}.

In addition to the central tracking detectors, the V0 detector is used for event characterization. The V0 detector is made of two plastic scintillator detectors, V0A and V0C, that cover the forward ($2.8<\eta<5.1$) and backward ($-3.7<\eta<-1.7$) pseudorapidity regions, respectively. The V0 detector is used to determine the charged-particle multiplicity and for triggering purposes.

\section{Data analysis}
\label{sec:data}
\subsection{Event selection}

The dataset analyzed in this paper consists of p--Pb collisions recorded in 2016 during LHC Run 2 at $\sNN{} =$ \SI{5.02}{\tera \eV}. Events were selected using the minimum bias (MB) trigger, which requires a hit in both the V0A and V0C detectors~\cite{V0_2013}. In addition, events are constrained to have their reconstructed primary vertex position within $|z_{\rm vtx}| < $~\SI{10}{cm} from the nominal interaction point in the beam-axis direction. To perform correlation measurements with a reconstructed $\phi$(1020) meson, only events with at least three charged tracks with $\pt{} > $~\SI{0.15}{\giga\eVperc} in the central $|\eta_{\text{lab}}| < 0.9$ region are used in this analysis. This requirement is also applied to the dihadron correlation measurement to ensure an accurate comparison.

 Due to the asymmetric energy-per-nucleon of the proton and lead-ion beams, the nucleon--nucleon center of mass is shifted by a rapidity of $\Delta y_{\text{NN}} = -0.465$ in the direction of the proton beam~\cite{phirapidity}. Events are  categorized based on their charged-particle multiplicity in the positive rapidity (incoming Pb beam direction) region as detected by the V0A, and are divided into multiplicity percentile intervals based on the multiplicity distribution of the entire dataset. For this analysis, three multiplicity ranges were chosen: 0--20\% (highest multiplicity), 20--50\%, and 50--80\% (lowest multiplicity). These intervals allow for grouping of events with similar event characteristics while still maintaining a significant $\phi$(1020) signal in each multiplicity range. After the trigger and offline event selection criteria are applied, a total of $\approx 4\times10^8$ events are selected for this analysis, corresponding to an integrated luminosity of $\approx 200 ~\SI{}{\micro\invbarn}$~\cite{luminosity_crosssection}. For simplicity, quoted rapidity and pseudorapidity ranges in this paper will refer to the laboratory frame.

\subsection{Track selection and \texorpdfstring{$\phi$}{Φ}(1020) reconstruction}
To measure the h--$\phi$ angular correlation function required for separating $\phi$(1020) production in jets and in the underlying event, and the h--h correlations to extract $\phi/\mathrm{h}$ ratios, three different types of particles are considered: high-momentum trigger particles (jet proxy),  $\phi$(1020) meson decay products, and inclusive charged hadrons for the dihadron correlation. Trigger particles are selected in the transverse momentum range of $4 < \pt{} <$ \SI{8}{\giga\eVperc}, while associated particles ($\phi$(1020) and charged hadrons) are split into lower ($1.5 < \pt{} <$ \SI{2.5}{\giga\eVperc}$)$ and upper ($2.5 < \pt{} <$ \SI{4}{\giga\eVperc}$)$ transverse momentum ranges. 

\subsubsection{Kaon identification}
For this analysis, the $\phi$(1020) meson is reconstructed through the decay channel $\phi(1020)~\rightarrow~\text{K}^{+}\text{K}^{-}$, and the yields are corrected for this decay branching ratio B.R. $= 0.491\pm0.005$~\cite{PDG:2022ynf}. Kaon candidates are selected from the central pseudorapidity region of $|\eta| < 0.8$. A standard set of quality selections is applied to ensure high quality for the kaon candidate tracks, as described in Ref.~\cite{Adam2016}. These include requirements on the minimum number of TPC clusters that compose the track ($N^{\text{clust}}_{\text{TPC}} \geq 80$ out of 159), as well as an upper limit on $\chi^2$ per TPC cluster in the track fitting ($\chi^2/N^{\text{clust}}_{\text{TPC}} \leq 4$). The tracks are also selected with a momentum-dependent requirement on the distance of closest approach (DCA) between the track and the reconstructed primary vertex, such that the DCA in the transverse plane ($\text{DCA}_{XY}$) is constrained by $\text{DCA}_{XY} < \bigl(0.0105+0.0350/(p_{\rm T}/(\SI{}{\giga\eVperc}))^{1.1}\bigr)$ \SI{}{cm}. This DCA requirement selects all tracks originating from within a radius $7\sigma_{XY}$ around the primary vertex in the $XY$ plane, where $\sigma_{XY}$ is the resolution of the distance of closest approach from the track to the vertex in the transverse plane~\cite{2014perf}. Since the $\phi$(1020) has a short lifetime ($\approx \SI{46}{fm/\text{\ensuremath{c}}}$)~\cite{PDG:2022ynf}, and therefore a decay vertex close to the primary vertex, this DCA selection removes particle tracks coming from secondary weak decays while still preserving the tracks of the $\phi$(1020) decay particles.

Kaon candidates for $\phi(1020)$ reconstruction are further selected using the PID information from both the TPC and TOF detectors~\cite{PhysRevC.101.044907,alicepPbstrange}. In the TPC, the energy loss (\dedx{}) of charged particles traveling through the detector gas can be calculated using the Bethe--Bloch formula, and the momentum-dependent \dedx{} signal of a candidate kaon can be compared with this expected value. Due to the finite resolution of the track momenta and energy loss in the TPC ($\sigma_{\text{TPC}} \approx 5\%$ of the \dedx{} signal~\cite{tpcperf}), the distribution of \dedx{} signals for real kaons at a given momentum is treated as a Gaussian with the mean at the expected kaon value, and a width $\sigma_{\text{TPC}}$ that is fit experimentally. Kaons are selected by requiring the difference between the measured and expected \dedx{} to be within $|(\dedx{}) - (\dedx{})_{\text{Bethe--Bloch}}|/\sigma_{\text{TPC}} < 3$. A similar process is followed for the timing signal measured by the TOF detector, selecting kaon candidates at a given momentum with the constraint $|\beta - \beta_{\text{K}}|/\sigma_{\text{TOF}}$, where $\beta_{\text{K}}$ is the expected track velocity given the kaon mass hypothesis and the specified momentum. These selections for PID allow for high efficiency in detecting produced kaons. The additional invariant mass reconstruction steps discussed in the following section eliminate all misidentified pions and protons from the set of candidate $\phi$ decay particles, as well as the combinatorial background due to uncorrelated kaon pairs.

\subsubsection{\label{sec:invmass}\texorpdfstring{$\phi$}{Φ}(1020) invariant mass reconstruction}
\begin{figure}[htb!]
\centering
\subfigure[][]{
\includegraphics[width=0.48\columnwidth]{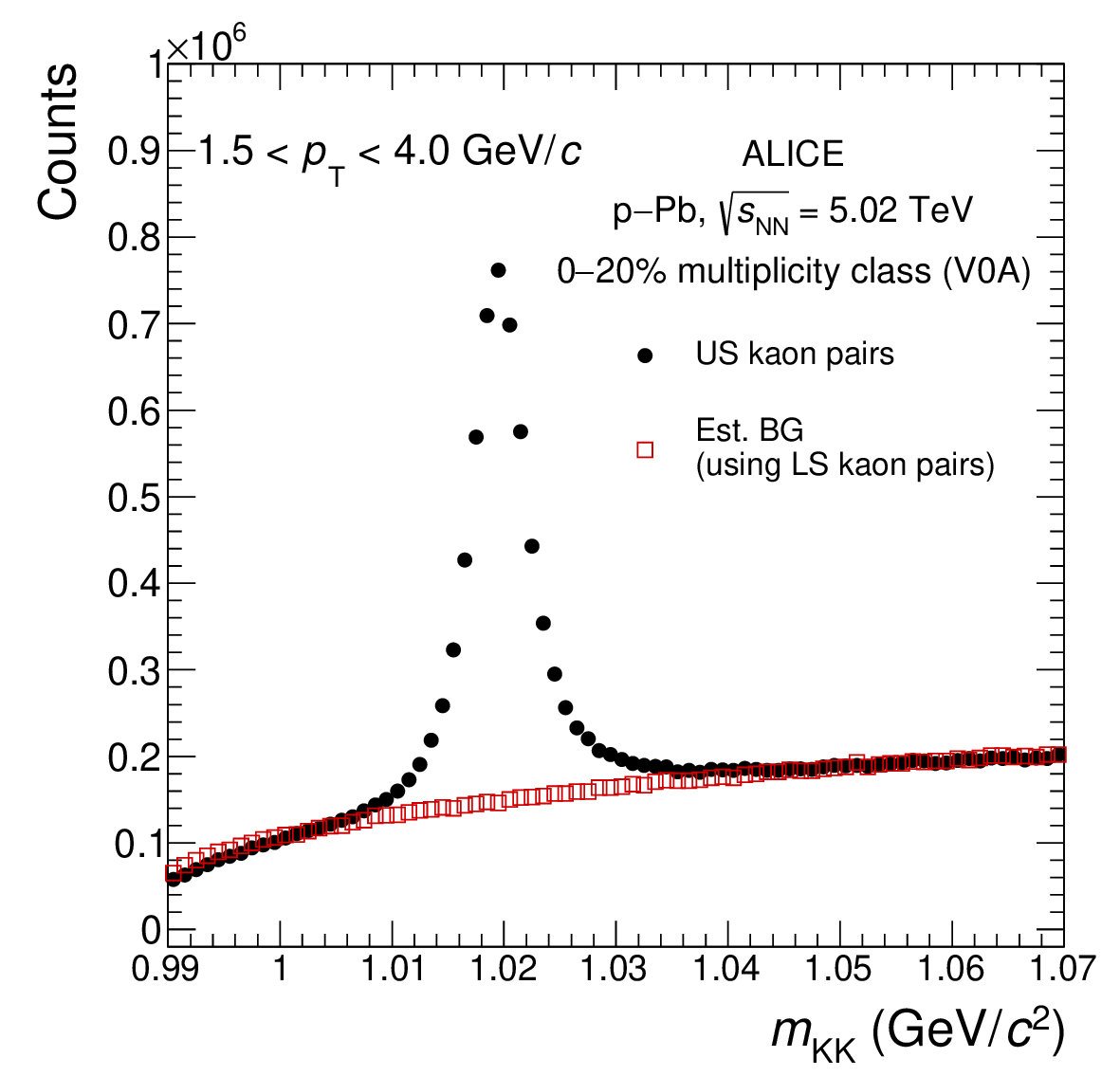}
\label{phiinvmass}
}
\subfigure[][]{
\includegraphics[width=0.48\columnwidth]{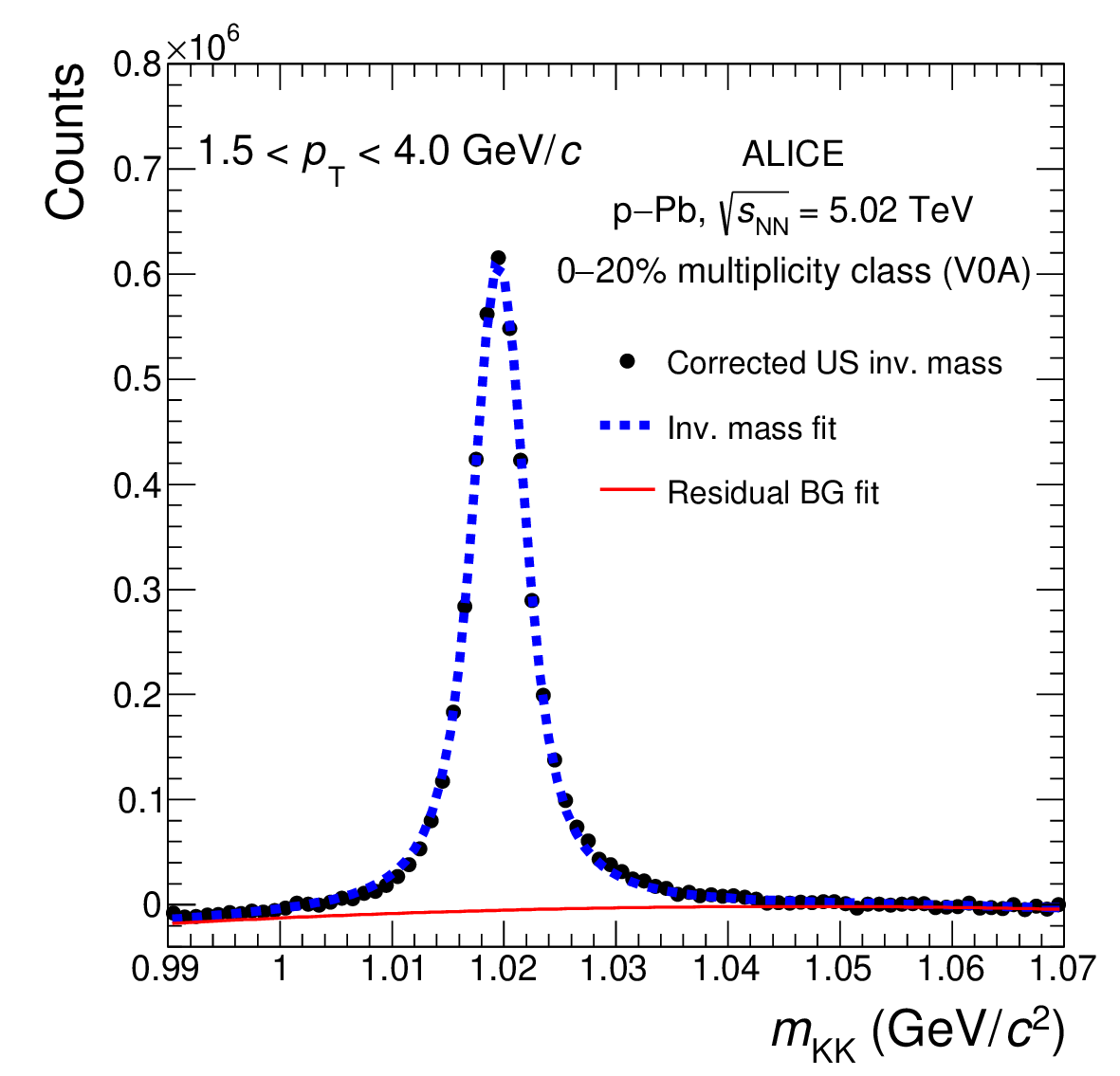}
\label{bgsubphiinvmass}
}
\caption{(a) Invariant-mass distribution of unlike-sign (US) kaon pairs (black), with the scaled like-sign (LS) pair distribution (red) used to approximate the combinatorial background (BG). (b) Background subtracted invariant-mass distribution of unlike-sign kaon pairs (black). The $\phi$(1020) mass peak is fit (blue dashed line) by a sum of a Voigt function and a first-order polynomial, which accounts for the residual background. This linear fit of the residual background is shown as a red line. Statistical uncertainties are plotted as bars, but are too small to be visible here.}
\end{figure}

 The invariant-mass distribution of the $\phi$(1020) meson is computed by considering all unlike-sign (US) pairs of kaons satisfying the selections described above, which includes both kaon pairs from $\phi(1020)$ decays as well as uncorrelated combinatorial kaon pairs. The invariant-mass distribution of like-sign (LS) kaon pairs, which is entirely composed of combinatorial kaon pairs, is used to estimate the combinatorial background underneath the mass-peak region of US kaon pairs (Fig.~\ref{phiinvmass}). For this purpose, the LS invariant-mass distribution is scaled to match the US distribution in the left and right sideband regions ($0.995 < m_{\text{KK}} < $ \SI{1.005}{\giga \eVperc \squared} and $1.04 < m_{\text{KK}} <$ \SI{1.06}{\giga\eVperc\squared}, respectively) away from the $\phi(1020)$ mass peak. The scale factor is defined as the sum of the yields in the left and right sidebands in the US mass distribution divided by the same yields in the LS mass distribution. The invariant mass distribution obtained after subtracting the scaled LS distribution is fitted with a Voigtian function (Gaussian convoluted with a relativistic Breit--Wigner distribution) for the resonance mass peak added to a polynomial term to describe the residual background, following the procedure described in Ref.~\cite{phiprod_ppPbPb}. For the \pt{} ranges considered in this analysis, the like-sign subtraction method leaves a residual background under the $\phi$(1020) mass peak of less than 1\% of the total signal as shown in Fig.~\ref{bgsubphiinvmass}.
 
 In order to ensure a large sample purity, the invariant-mass region used to extract the $\phi$(1020) yield in this analysis is defined as $\pm$ \SI{6}{\mega\eVperc^2} around the nominal $\phi(1020)$ mass (i.e., \mbox{$1.014 < m_{\mathrm{KK}} < $ \SI{1.026}{\giga \eVperc \squared}}). However, this region does not encompass the entire mass distribution of the signal, and therefore an additional factor is needed to correct for the missing distribution tails while calculating the yield. This factor, denoted as $k_{\text{Signal}}$, is computed as the integral of the Voigtian fit across the entire mass range divided by the integral in the selected mass-peak region, resulting in a correction factor $k_{\text{Signal}} = 1.22$.

 To correct for the finite acceptance and inefficiencies of the detector, a Monte Carlo simulation of minimum-bias events based on the PYTHIA 8 event generator~\cite{pythia} is used to estimate the acceptance $\times$ efficiency ($A \times \varepsilon_{\rm rec}$) for the $\phi$(1020). Within the simulated events, the $\phi$(1020) decay products are propagated through a full simulation of the ALICE detector material using GEANT3~\cite{geant}. The $A\times\varepsilon_{\rm rec}$ of the $\phi$(1020) is then defined as the number of $\phi$(1020) at mid-pseudorapidity ($|\eta_{\phi}| < 0.8$) that are reconstructed after the decay products pass the track quality, PID, and pseudorapidity requirements ($|\eta_{\mathrm{track}}| < 0.8$) divided by the total number of $\phi$(1020) that were produced in the same pseudorapidity window in the simulated events. With the track quality selections described above, the two-track reconstruction efficiency fraction increases from 0.3 to 0.6 with increasing \pt{} in the range considered in this analysis ($1.5 < p^{\phi}_{\rm T} <$ \SI{4.0}{\giga\eVperc}). The PID selection and the requirement for a matching TOF hit introduces another factor ranging from 0.25 to 0.4, leading to a final $A\times\varepsilon_{\rm rec}$ for the $\phi$(1020) ranging from 0.09 to 0.22. This efficiency is independent of  multiplicity, and is therefore calculated from the full minimum-bias sample of events. 

\subsection{Associated hadrons and trigger track selection}

For the associated hadrons in the dihadron correlation measurement, the same track-quality selections are used that are required for the kaon candidates ($N^{\text{clust}}_{\text{TPC}} \geq 80$, $\chi^2/N^{\text{clus}}_{\text{TPC}} \leq 4$, $|\eta| < 0.8$, the \pt{}-dependent DCA requirement). These selection criteria are used to select associated hadrons that are physical primary particles, defined as those hadrons that are produced either directly from the collision or from resonance decays~\cite{ALICEprimary}. After primary selection, secondary decay products contribute less than $1\%$ to the total associated hadron yield. The tracking efficiency for the associated charged hadrons is approximately 0.8 for the entire \pt{} range considered of $1.5 < \pt{} <$ \SI{4}{\giga\eVperc}.

For the high-\pt~trigger hadron, tracks are selected in the transverse momentum range $4 < \pt{} <$ \SI{8}{\giga\eVperc}, and a slightly modified set of track quality requirements is used in order to achieve higher track reconstruction efficiency. These selections, also detailed in Ref.~\cite{jetlikecorr}, relax the \pt-dependent DCA requirement and replace it with a simple \SI{2.4}{cm} (\SI{3.2}{cm}) cutoff in the transverse (longitudinal) distance to the primary vertex. This gives a reconstruction acceptance times efficiency ($A\times\varepsilon_{\rm rec}$) of approximately 0.85 for the trigger hadron. 

The validity of the above efficiency corrections is checked using Monte Carlo closure tests, where the fully efficiency corrected h--$\phi$ and h--h measurements are built from reconstructed tracks in PYTHIA 8 generated MC events. The reconstructed MC distributions are in good agreement with the generated MC h--$\phi$ and h--h signal.

\section{Angular correlation measurement}
\label{sec:corr}
A two-particle correlation function in two-dimensional angular space can be written as:
\begin{equation}
     C(\varphi_t, \eta_t, \varphi_a, \eta_a) = \frac{P_{t,a}(\varphi_t, \eta_t, \varphi_a, \eta_a)}{P_t(\varphi_t, \eta_t)P_a(\varphi_a, \eta_a)},
\end{equation}
where $P_{t,a}$ is the joint probability distribution of finding an associated particle at azimuthal angle $\varphi_a$ and pseudorapidity $\eta_a$ along with a trigger particle at azimuthal angle $\varphi_t$ and pseudorapidity $\eta_t$, and $P_t$ and $P_a$ are the single-particle probability distributions. This correlation function can be simplified experimentally by using only the measured relative angles $\Delta\varphi$ and $\Delta\eta$ between the trigger and associated particles. Using this simplification, an experimental correlation function can be approximated as:
\begin{equation}
    C(\Delta\varphi, \Delta\eta) \approx \frac{S(\Delta\varphi, \Delta\eta)}{B(\Delta\varphi, \Delta\eta)},
\end{equation}
where $S(\Delta\varphi, \Delta\eta) = \frac{{\rm d}^2 N_{\text{pair}}}{{\rm d}\Delta\varphi {\rm d}\Delta\eta}$ is the distribution of trigger--associated particle pairs from the same event, and $B(\Delta\varphi, \Delta\eta)$ is the mixed-event distribution, constructed by taking trigger and associated particles from different events. The mixed-event distribution depends on the single-particle distributions, and represents the two-particle acceptance of trigger--associated pairs~\cite{finitecorr}. This correlation function can be scaled to give the per-trigger yield of associated particles:
\begin{linenomath}
\begin{equation}
    C_{\rm trig}(\Delta\varphi, \Delta\eta) = \frac{1}{N_{\rm trig}}\frac{{\rm d}^2 N_{\text{pair}}}{{\rm d}\Delta\varphi {\rm d}\Delta\eta} =  \frac{1}{N_{\rm trig}^{\rm corr}}\frac{1}{\varepsilon_{\rm trig}\times \varepsilon_{\rm assoc}}\frac{B(0,0)\times S(\Delta\varphi, \Delta\eta)}{B(\Delta\varphi, \Delta\eta)},
\end{equation}
\end{linenomath}
where $N_{\rm trig}^{\rm corr}$ is the efficiency corrected total number of triggers, the $\varepsilon_{\rm trig}\times \varepsilon_{\rm assoc}$ denotes the product of the \pt-dependent acceptances and efficiencies for the trigger hadron and associated particle, and $S(\Delta\varphi, \Delta\eta)$ and $B(\Delta\varphi, \Delta\eta)$ are the same-event and mixed-event correlation distributions, respectively~\cite{phenixcorr, cmscorr, othercmscorr}. Both the same- and mixed-event correlations are corrected by the trigger and associated particle efficiency. Here the mixed-event correlation is used as a pair acceptance correction, and is normalized by the factor $B(0,0)$. This ensures that trigger and associated particles traveling in the same direction ($\Delta\varphi\approx 0, \Delta\eta\approx 0$) have a pair acceptance equal to 1~\cite{alicePbPbcorr}. This way of restructuring the correlation function gives the number of correlated trigger--associated particle pairs per trigger, or the yield of associated particles per trigger.

When performing the correlation with a $\phi$(1020) meson, it is necessary to remove the combinatorial background from the pool of $\phi$(1020) candidates. To do this, the correlation measurement is performed for three different US kaon-pair invariant mass regions: the $\phi$(1020) mass-peak region ($C_{\rm trig}^{\text{h--(KK)}\text{ Peak}}$), the left sideband region ($C_{\rm trig}^{\text{h--(KK)}\text{ LSB}}$), and the right sideband region ($C_{\rm trig}^{\text{h--(KK)}\text{ RSB}}$). The correlations in the two sideband regions are averaged together and used to estimate the shape of the distribution produced by correlating the trigger hadron with a combinatorial kaon pair. Taking into account these additional terms, the final correlation function is given as:
\begin{linenomath}
\begin{align}
\label{fullCorrEq}
\begin{split}
    C_{\rm trig}^{\text{h--}\phi}(\Delta\varphi, \Delta\eta)& = k_{\text{Signal}}\biggl(C_{\rm trig}^{\text{h--(KK)}\text{ Peak}}(\Delta\varphi, \Delta\eta)\\
    &- \frac{k_{\text{LS}}}{2}\biggl[\frac{1}{N_{\rm Int}^{\rm LSB}}C_{\rm trig}^{\text{h--(KK)}\text{ LSB}}(\Delta\varphi, \Delta\eta) + \frac{1}{N_{\rm Int}^{\rm RSB}}C_{\rm trig}^{\text{h--(KK)}\text{ RSB}}(\Delta\varphi, \Delta\eta) \biggr]\biggr),
\end{split}
\end{align}
\end{linenomath}
where the factor $k_{\text{LS}}$ is the integral of the scaled LS distribution in the mass-peak region, $1/k_{\text{Signal}}$ is the fraction of $\phi(1020)$ in the mass-peak region (defined in Section~\ref{sec:invmass}), and factors $N_{\rm Int}^{\rm RSB}$ and $N_{\rm Int}^{\rm LSB}$ are defined as the total integral of the correlations for the right and left sideband regions, respectively. Scaling by the inverse of these total integral factors normalizes the correlation functions in the sideband regions to an integral of 1, before scaling up by factor $k_{\text{LS}}$ to estimate the background under the peak region.

Once the correlation is computed, the angular correlation is divided up into three different regions to calculate the yields of correlated pairs: the near-side jet (centered at $\Delta\varphi=0$), the away-side jet (centered at $\Delta\varphi=\pi$), and the underlying event which lies under both jet peaks. To reduce the error in this separation due to statistical noise, the two-dimensional correlation is projected onto $\Delta\varphi$ from the range $-1.2 < \Delta\eta < 1.2$. The range for $\Delta\varphi$ is chosen to be $[-\pi/2, 3\pi/2]$. In order to separate the jet peaks from the underlying event in this one-dimensional correlation, a flat background estimate is used by computing the average over the region farthest away from the two jet peaks ($-\pi/2 < \Delta\varphi < -\pi/4$, $\pi/4 < \Delta\varphi < 5\pi/8$, and $11\pi/8 < \Delta\varphi < 3\pi/2$). The near-side (away-side) yield is then defined as the area above this background in the region $-\pi/2 < \Delta\varphi < \pi/2$ ($\pi/2 < \Delta\varphi < 3\pi/2$). The underlying-event yield is computed as the area underneath this flat background across the full $\Delta\varphi$ range.

\section{Systematic uncertainties} 
\label{sec:syst}
The dominant systematic uncertainties for this analysis are divided into two main categories: those associated with the h--$\phi$ correlation measurement directly, which are related to the detector tracking efficiency, the identification of produced kaons, and the $\phi(1020)$ reconstruction method; and those associated with separating the correlation measurement into jet and underlying-event contributions. The tracking systematic uncertainties include contributions from the global tracking efficiency, the material budget of the detector, and the track selection cuts. These uncertainties are estimated as in Ref.~\cite{pPbproduction}, and are found to be \pt{} independent. Their contribution provides the largest uncertainty to the associated yields, approximately 3.6\% for the hadron yields and 5.2\% for the yields of the reconstructed $\phi$(1020). The error associated with the PID selection of kaon candidates was estimated by independently varying the ranges for $n\sigma_{\mathrm{TPC}}$ and $n\sigma_{\mathrm{TOF}}$ between $n=2$ and $n=3$ by steps of 0.2 and 0.5 for the TPC and TOF selection cuts, respectively. 

For the h--$\phi$ correlation, additional systematic uncertainties due to the $\phi(1020)$ reconstruction method are estimated by varying the invariant-mass ranges used for the $\phi$(1020) mass peak, and varying the left and right sideband mass regions used for combinatorial background subtraction. The mass-peak region is varied from $1.010 < m_{\mathrm{KK}} < \SI{1.030}{\giga \eVperc^2}$ to $1.015 < m_{\mathrm{KK}} < \SI{1.025}{\giga \eVperc^2}$ in steps of $\SI{2}{\mega\eVperc^2}$, leading to a variation of the total fraction of $\phi$(1020) signal that is measured from 90\% to 76\%. The systematics due to the choice of sideband regions are estimated by independently varying the upper and lower limit of the left and right sideband by $\pm\SI{5}{\mega\eVperc^2}$, as well as using only the left or only the right sideband for the background estimation. The contributions of these systematic uncertainties are detailed in Table~\ref{systtable}. All systematic effects considered for the correlation measurement are $\Delta\varphi$ independent and are applied evenly to the full $\Delta\varphi$ correlation. These uncertainties also do not vary within the \pt{} ranges considered, and therefore are applied equally to the two \pt{} ranges considered in this analysis. The dominant systematic uncertainty for the correlation emerges due to uncertainties in the tracking efficiency correction.

\begin{table}[htb!]
    \centering
    \caption{Summary of the systematic uncertainty on the h--$\phi$ angular correlation for each multiplicity interval, expressed in percent of the central value. The table also quotes the average statistical relative error per $\Delta\varphi$ interval for comparison.}
\begin{tabular}{ c | c | c | c }
	\multicolumn{4}{c}{h--$\phi$ Relative Error per $\Delta\varphi$ Bin} \\
 \hline
	Multiplicity & 0--20\% &	20--50\%	& 50--80\% \\
 \hline
Tracking eff. &	5.2\% &	5.2\% &	5.2\% \\
PID selection	& 1.0\%	& 1.0\% & 1.0\% \\
Mass-peak range	& 0.8\%	 & 1.3\% & 2.7\% \\
LSB mass range&	1.1\%	&0.9\%	&2.6\% \\
RSB mass range&	0.2\% &	0.3\% &	1.2\% \\
$k_{\text{LS}}$ scale factor&	0.9\% &	0.6\% &	3.0\% \\
$k_{\text{Signal}}$ scale factor &	0.8\% &	0.8\% &	0.8\% \\
\hline
Total syst.	& 5.6\%	& 5.6\%	& 7.3\% \\
Avg. statistical error & 2.2\% & 3.2\% & 7.3\% \\
\end{tabular}

\label{systtable}
\end{table}

In addition to the sources of systematic uncertainty for the correlation measurement in Table~\ref{systtable}, the pairwise yield measurement includes an additional uncertainty stemming from the method of determining the underlying-event baseline to separate the jet and non-jet yields. This systematic is determined using a variety of different background estimation methods:  fitting a constant to different $\Delta\varphi$ subranges around the minima of the correlation function (i.e., only considering points around the near-side peak, only points around the away-side peak, and combinations of both subranges), the zero yield at minimum (ZYAM) method as detailed in Refs.~\cite{PhysRevLett.97.052301,PhysRevLett.90.082302}, and a full fit with two Gaussian distributions for the near- and away-side peaks plus a constant for the background. The systematic uncertainty on the background is determined by taking the standard deviation of these values across the different background methods, giving an uncertainty for the near-side yields ranging from $\approx$2\% in the high-multiplicity h--$\phi$ correlation, to $\approx$5\% in the lowest multiplicity class. The h--$\phi$ away-side yields, being more dependent on the background estimation, are found to have a systematic uncertainty of $\approx$7\% for each multiplicity class, while the effect on the underlying event yields was negligible. For the h--h correlations, all different background estimation methods give similar values ($< 1\%$ difference to the nominal yields), and the systematic effects from the h--h background estimation are considered as negligible. For both the h--$\phi$ and h--h yield measurements, the systematic uncertainties are correlated between the near- and away-side yields, and therefore have a negligible effect on differences between the two.

The central measurement uses a flat background estimate to separate jet and underlying-event contributions. However, previous studies have measured collective behavior of particles produced within p--Pb collisions~\cite{smallsystem_flow}. In two-particle angular correlations, this collective expansion manifests itself as an anisotropic azimuthal distribution, which can be expanded as a Fourier series in $\Delta\varphi$. Higher-order components of the collective flow have a negligible effect at the momentum ranges studied here. However, the second Fourier coefficient ($v_2$), referred to as elliptic flow and characterized by the simplified expression $a_0(1 + 2 v^{\text{h}-\phi}_2\cos(2\Delta\varphi))$, does affect our yield measurements. Given the absence of published direct measurement of the $\phi$(1020) meson $v^{\phi}_2$ in p--Pb collisions, we have chosen the flat background assumption as the basis for the central values and include an additional systematic uncertainty for the near- and away-side jet yields based on using the charged hadron $v^{\text{h}}_2$ measured by the ALICE Collaboration~\cite{pPbv2}. Recent studies have also measured a non-zero third-order Fourier coefficient ($v_3$) in small systems~\cite{Acharya2024}. However, the effect on the correlation measurement due to the measured $v^{\text{h}}_3$ is a factor of 10 smaller than for the $v^{\text{h}}_2$ in the momentum ranges considered here, and is therefore not included in this analysis. Cross-checks performed using long-range correlations to estimate the $v^{\phi}_2$ from the p--Pb data being used for this paper are consistent with the inclusive charged-hadron $v^{\text{h}}_2$ that was chosen as an estimate. As the measured $v^{\text{h}}_2$ is momentum dependent, this cross-check is performed separately for the lower associated particle \pt{} interval ($1.5<\pt{}<$~\SI{2.5}{\giga\eVperc}) and the higher associated particle \pt{} interval ($2.5<\pt{}<$~\SI{4.0}{\giga\eVperc}). For the low momentum h--$\phi$ (h--h) yields, the non-zero $v_2$ assumption lowers the jet yields between 4\% (1\%) for the lowest multiplicity and 25\% (16\%) for the highest multiplicity. Likewise, the high momentum h--$\phi$ (h--h) yields are lowered between 1.4\% (0.6\%) for the lowest multiplicity and 16\% (9\%) for the highest multiplicity. Due to the non-zero $v_2$ affecting the near- and away-side yields in similar amounts, this phenomenon does not affect any differences in behavior between the two jet peaks. Likewise, because the h--$\phi$ and h--h correlations are affected by the $v_2$ in similar amounts, the impact on the final ratio is small, with a negligible effect at low multiplicity and an approximately 10\% (8\%) effect on the highest-multiplicity jet-yield ratios for the measurement in the low (high) associated \pt{} interval.

\section{Results}
\label{sec:results}
\begin{figure}[htb!]
\centering
\includegraphics[width=0.31\linewidth]{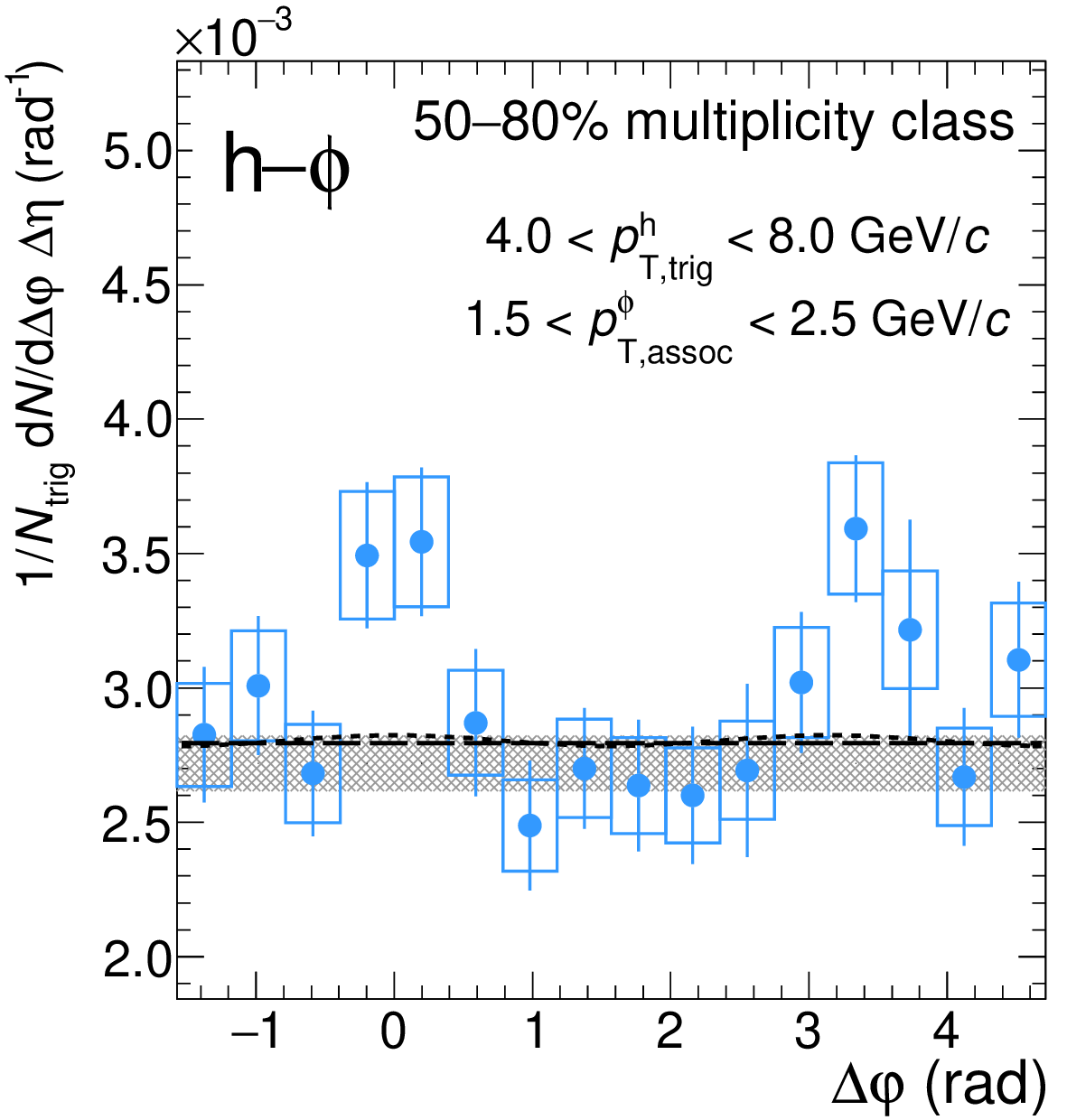}
\includegraphics[width=0.31\linewidth]{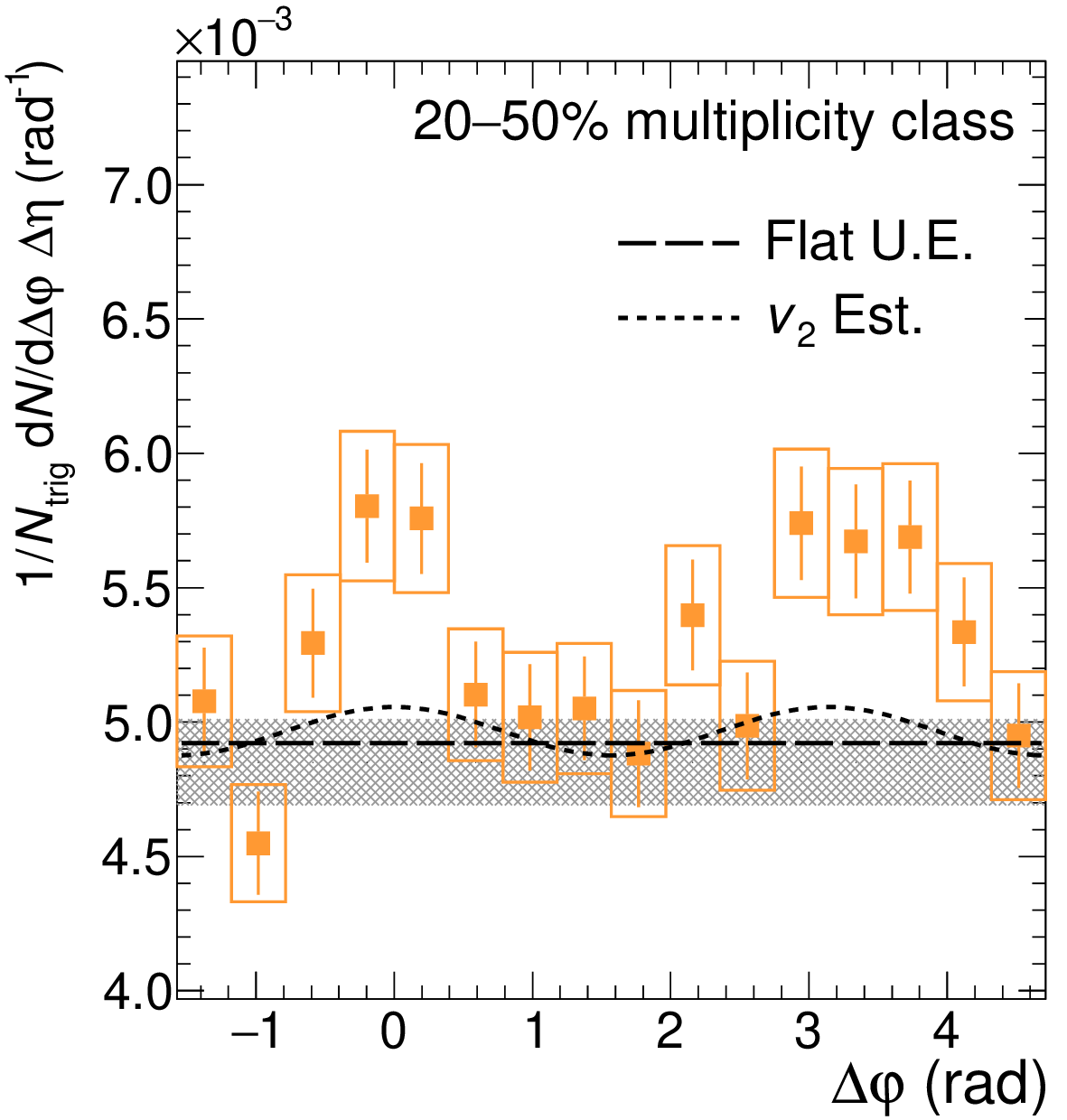}
\includegraphics[width=0.31\linewidth]{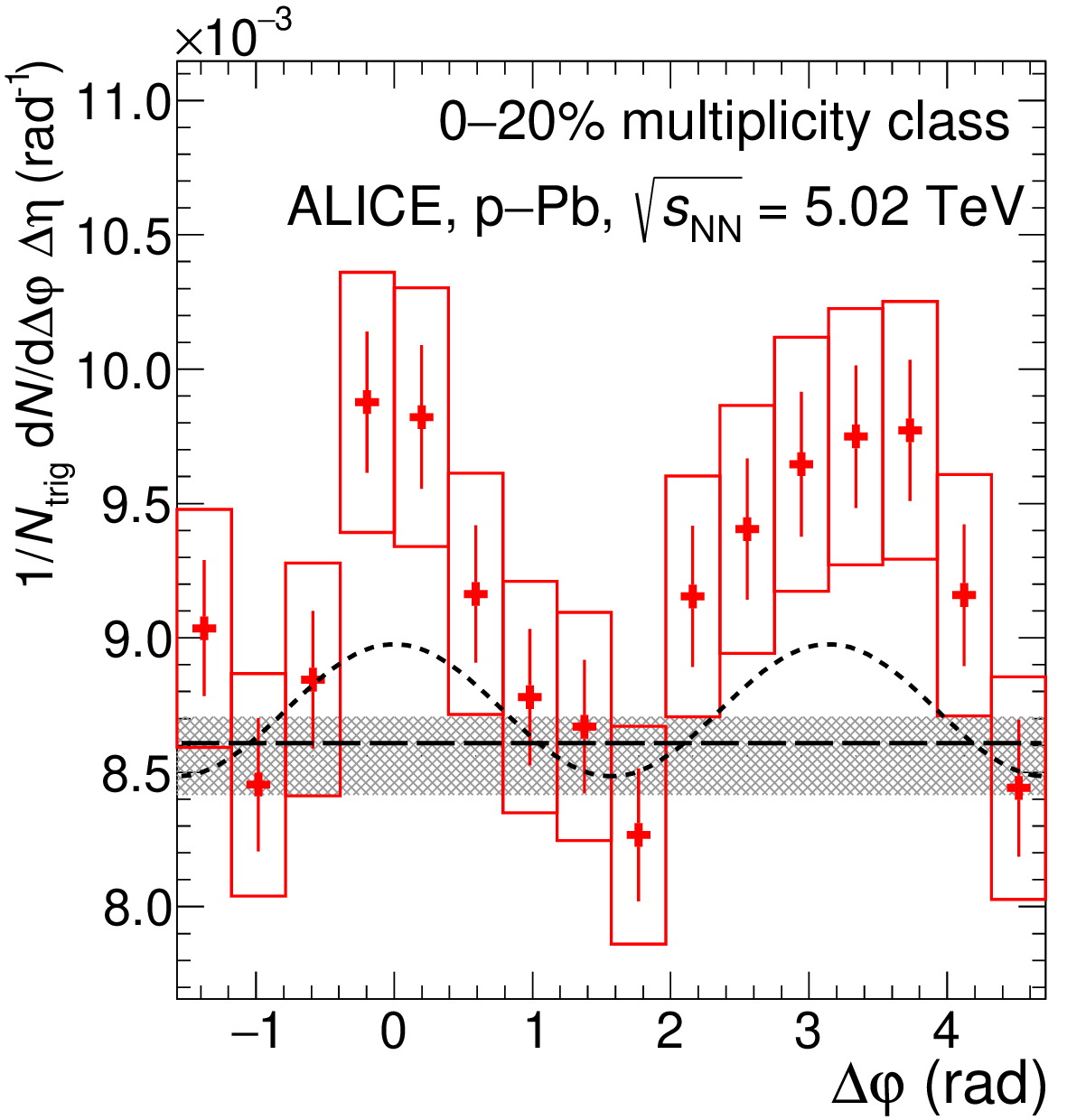}
\includegraphics[width=0.31\linewidth]{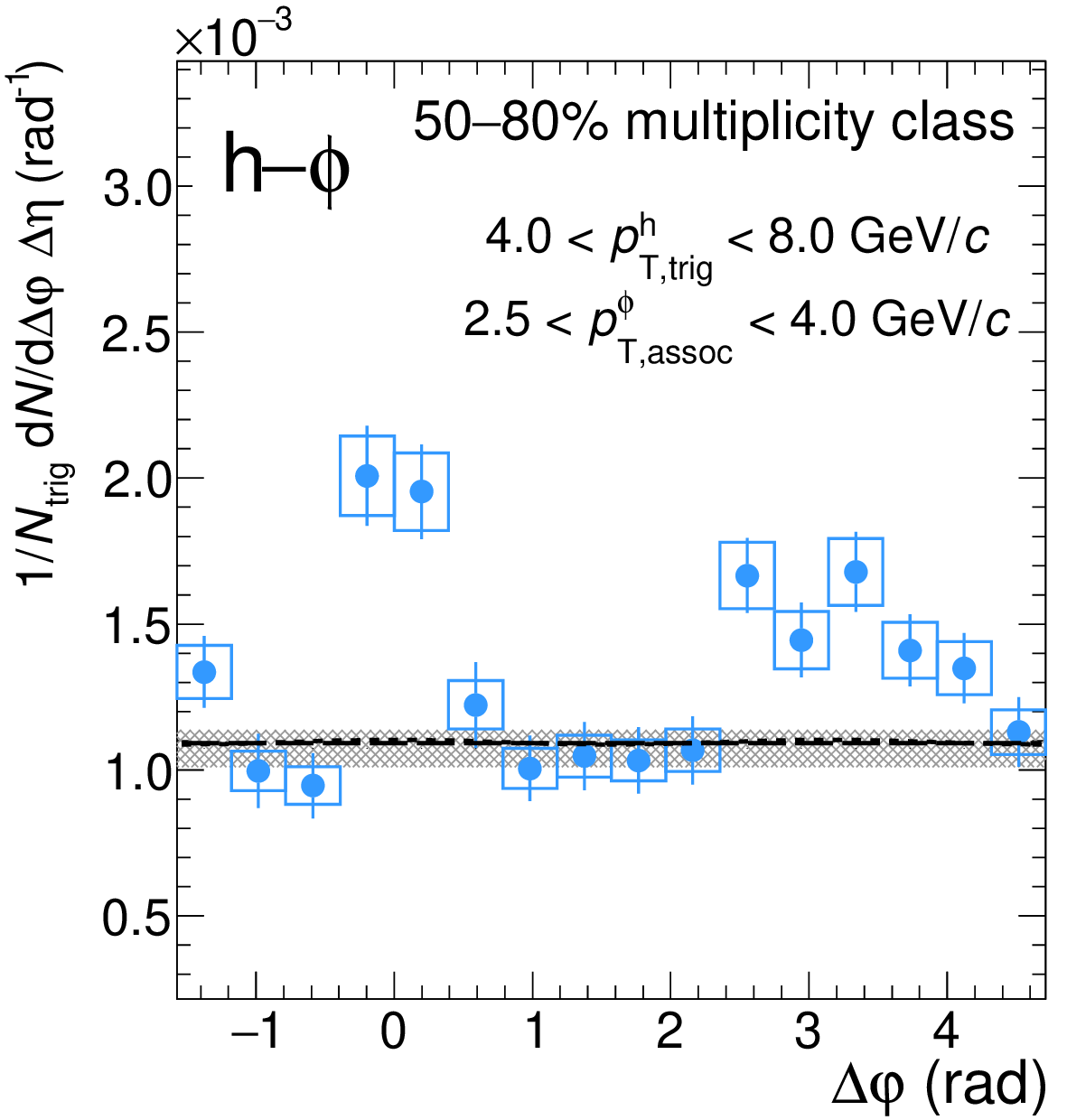}
\includegraphics[width=0.31\linewidth]{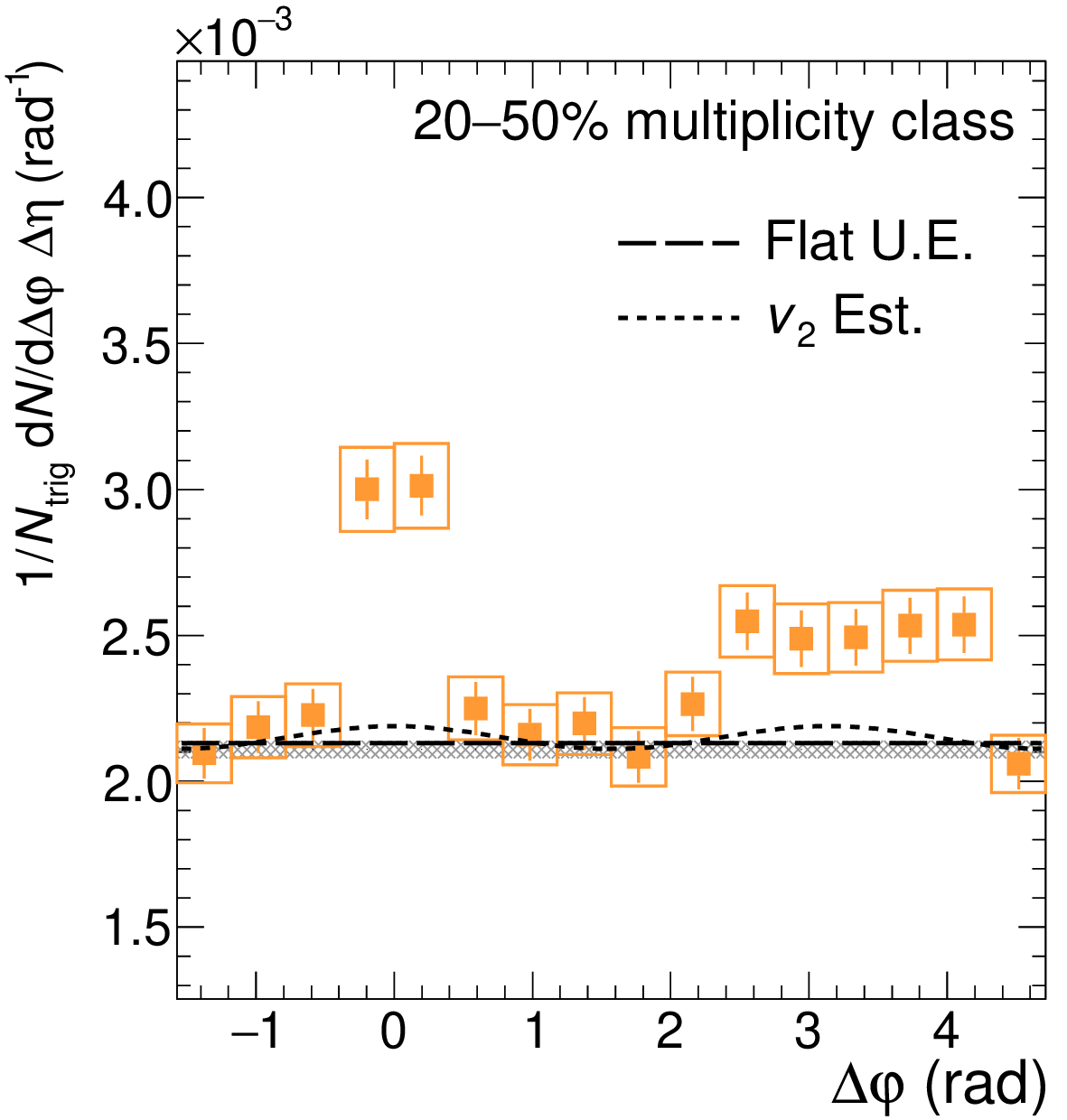}
\includegraphics[width=0.31\linewidth]{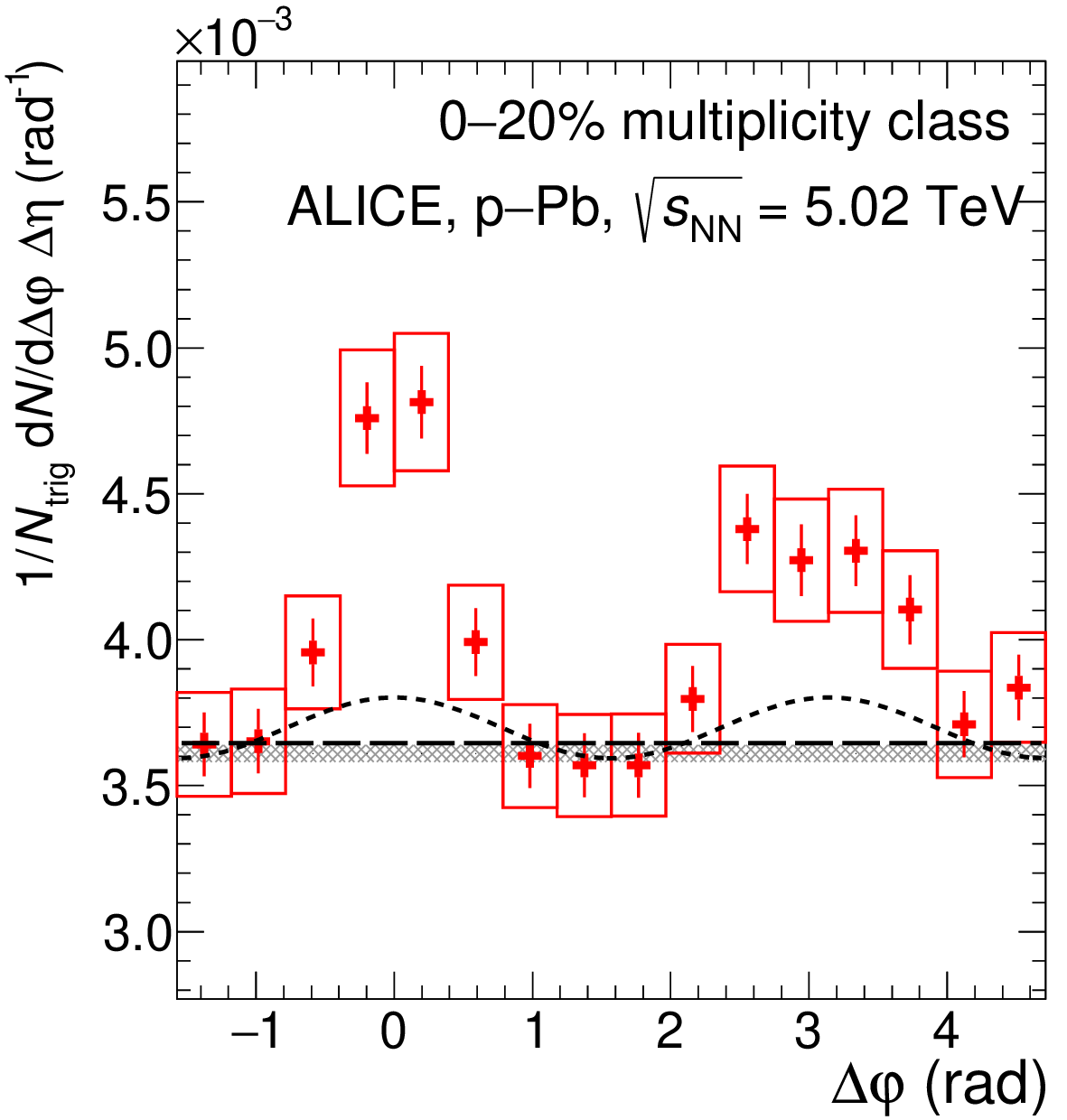}
\caption{Per-trigger h--$\phi$ $\Delta\varphi$ angular correlations integrated over the range $|\Delta\eta| < 1.2$ for the lower \pt{} range of associated $\phi$(1020), $1.5 < p_{\rm T} < $~\SI{2.5}{\giga\eVperc} (top), and the higher \pt{} range of associated $\phi$(1020), $2.5 < p_{\rm T} < $~\SI{4.0}{\giga\eVperc} (bottom). The three multiplicity classes are plotted in increasing order from left to right. All plots for a given associated particle \pt{} have the same axis scale, with offsets set to their respective underlying background estimation. Statistical uncertainties are represented as error bars, while systematic uncertainties are marked with open rectangles. The grey band shows the systematic range for the different background estimation methods. The long dashed line shows the flat background assumption, while the short dotted line shows the alternative background with non-zero $v_2$ (used as systematic for pair yields).}
\label{dphi_hphi}
\end{figure}

\begin{figure}[htb!]
\centering
\includegraphics[width=0.31\linewidth]{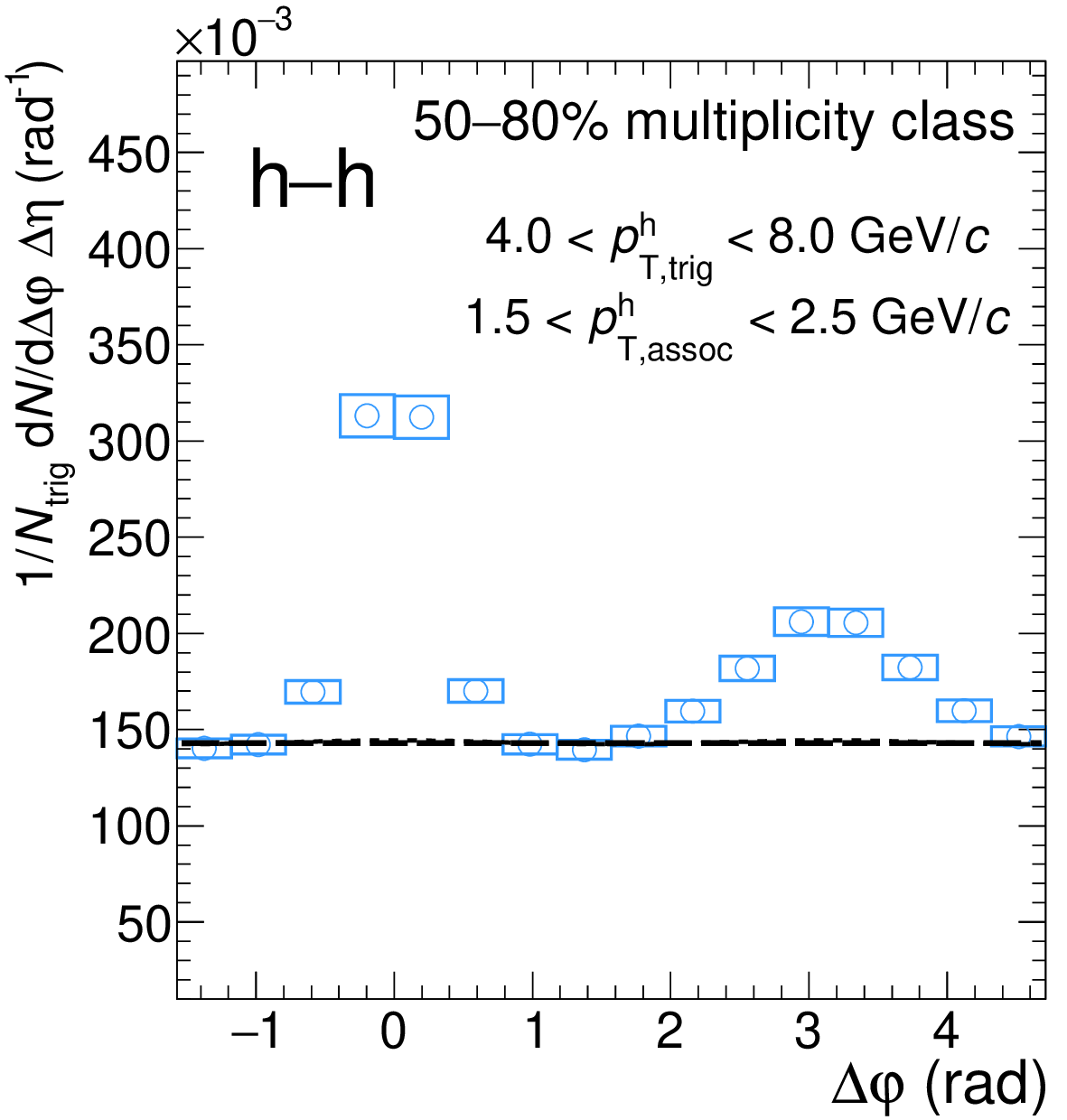}
\includegraphics[width=0.31\linewidth]{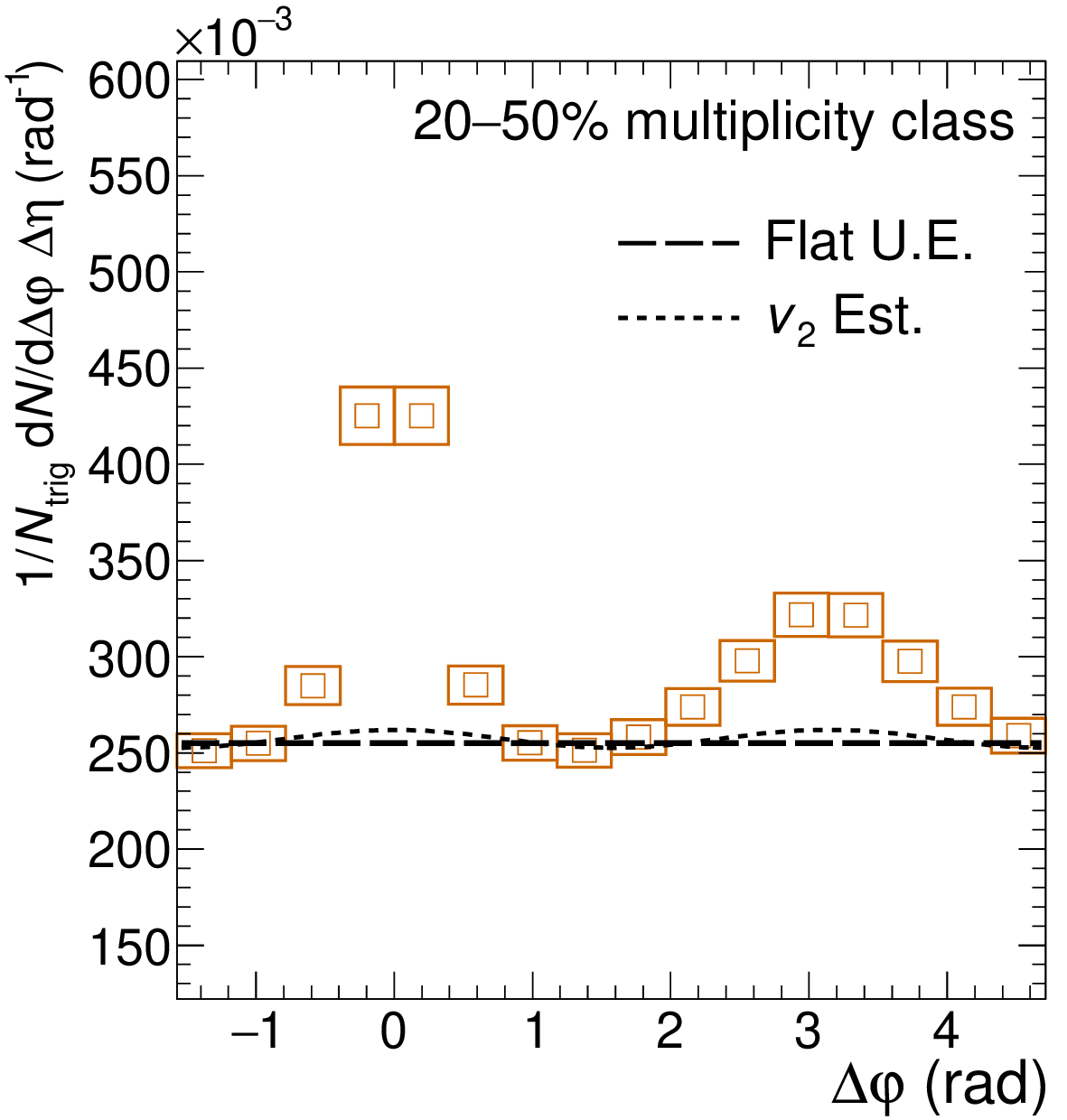}
\includegraphics[width=0.31\linewidth]{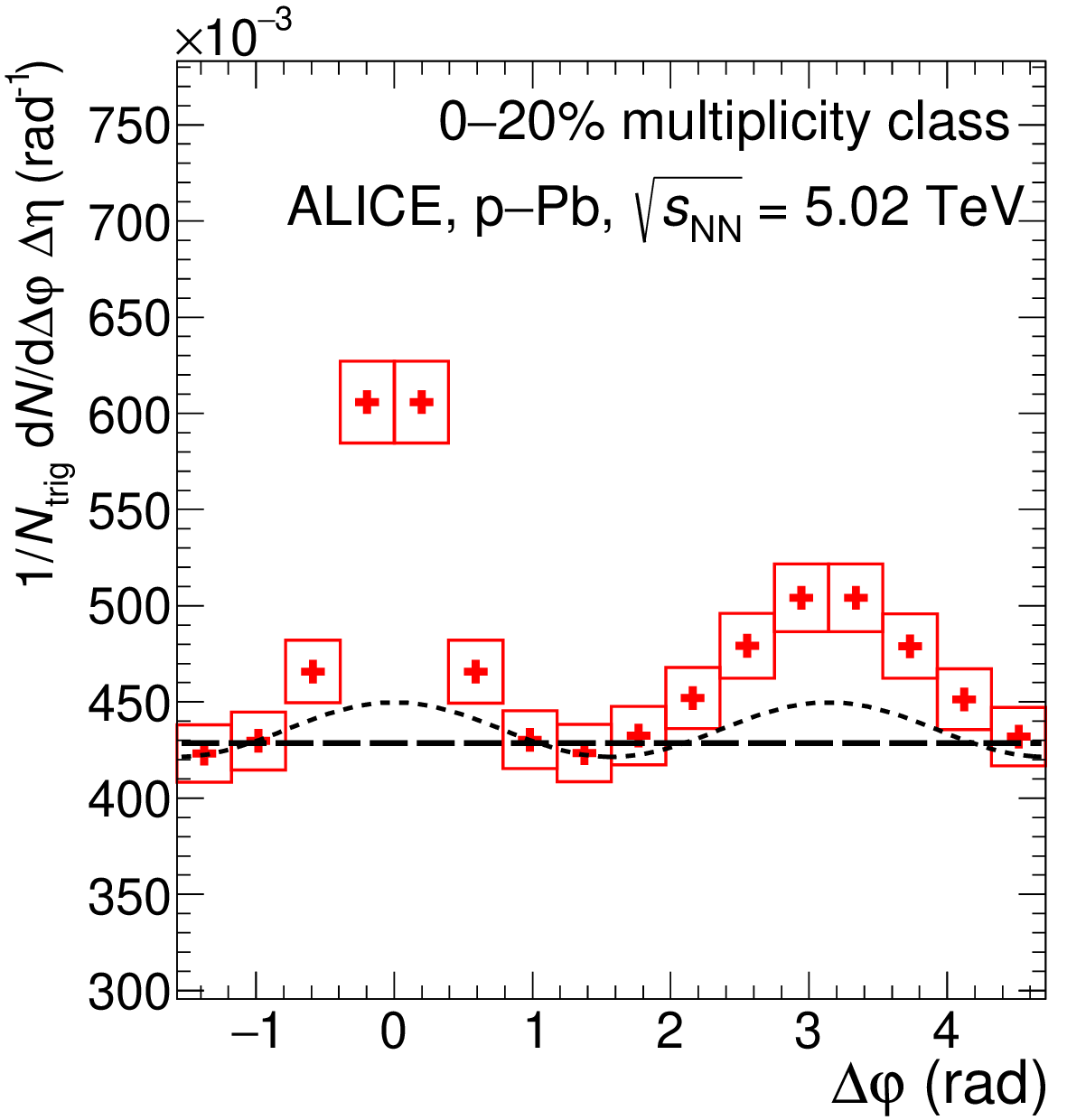}
\includegraphics[width=0.31\linewidth]{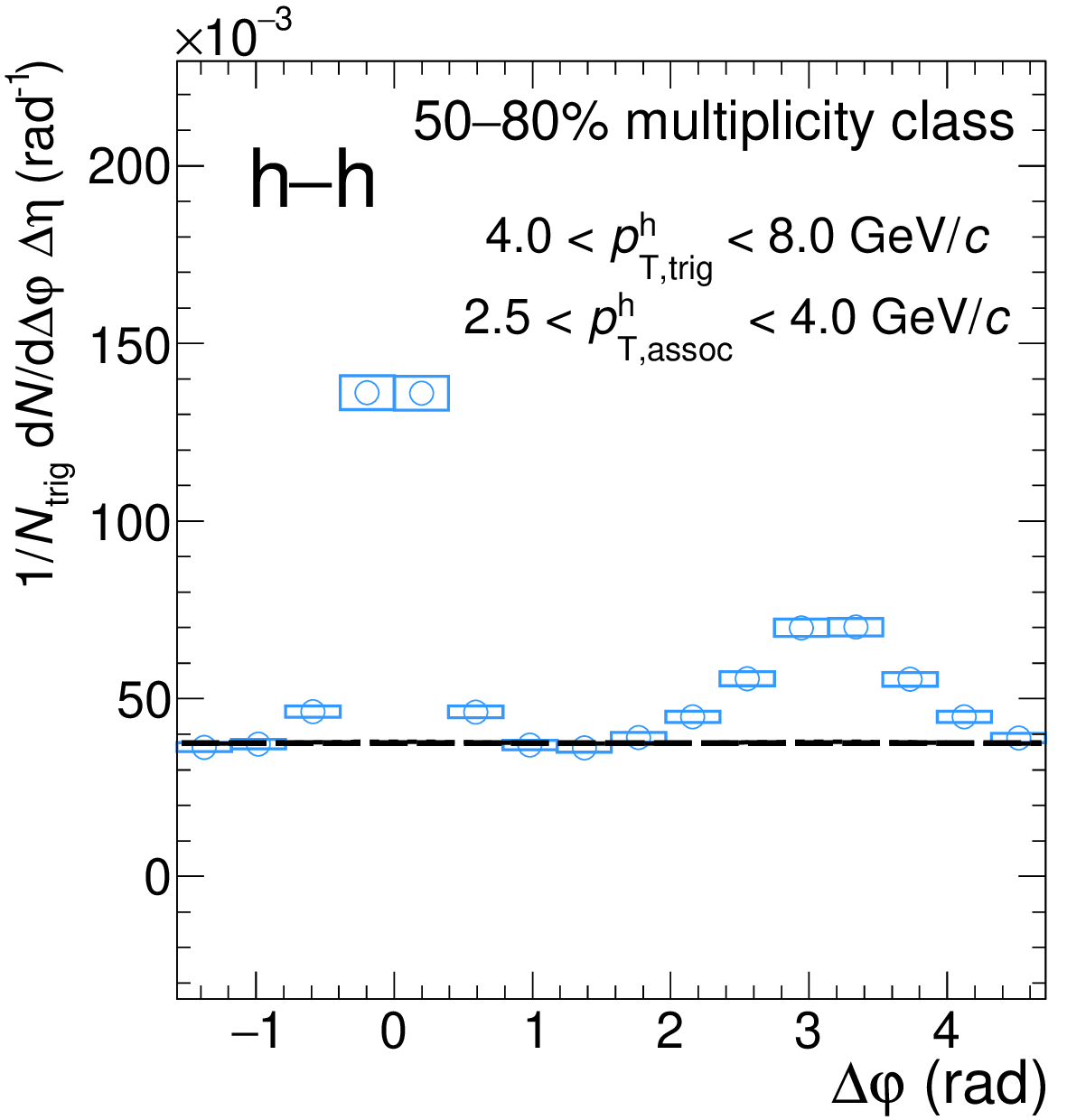}
\includegraphics[width=0.31\linewidth]{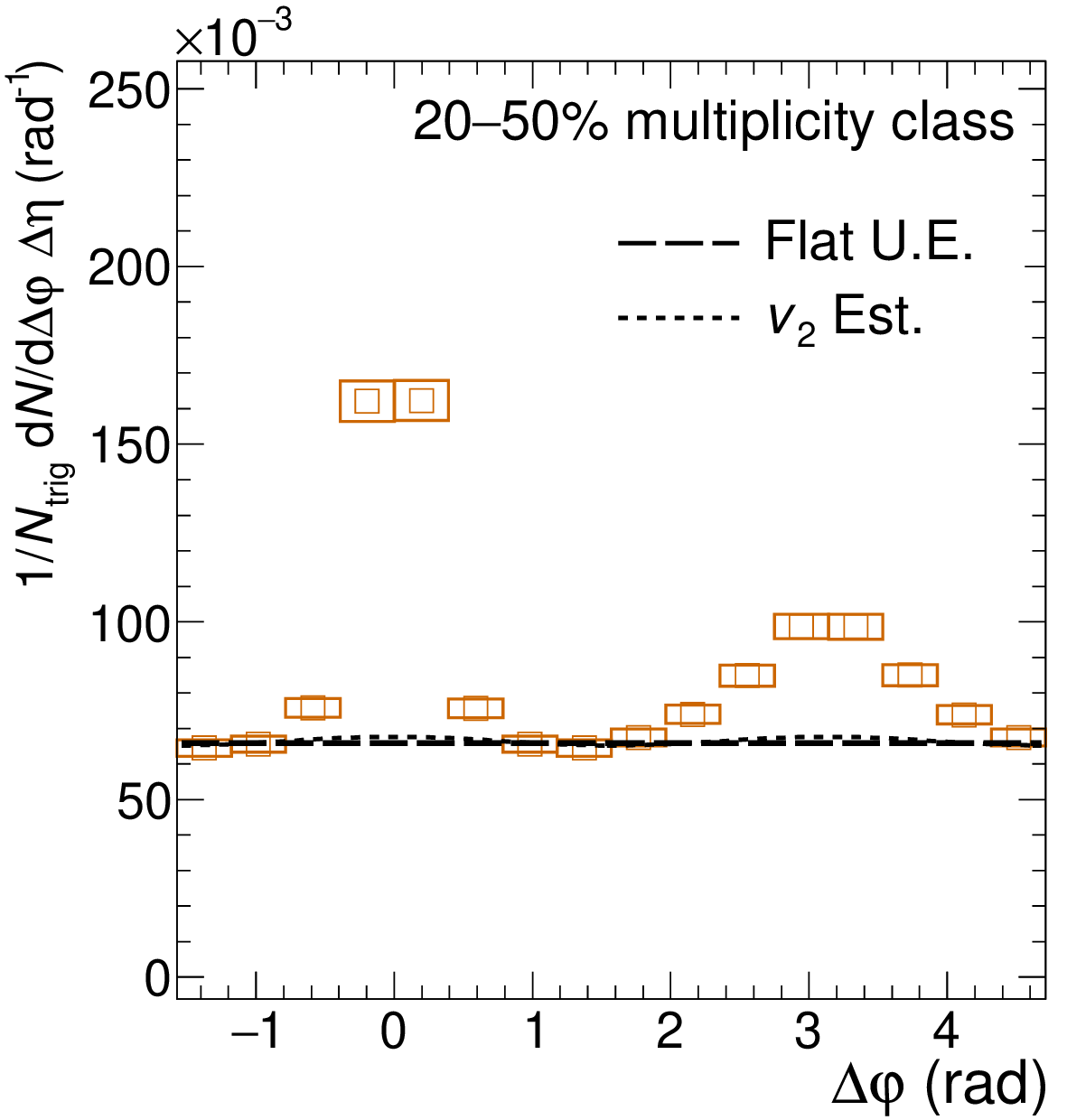}
\includegraphics[width=0.31\linewidth]{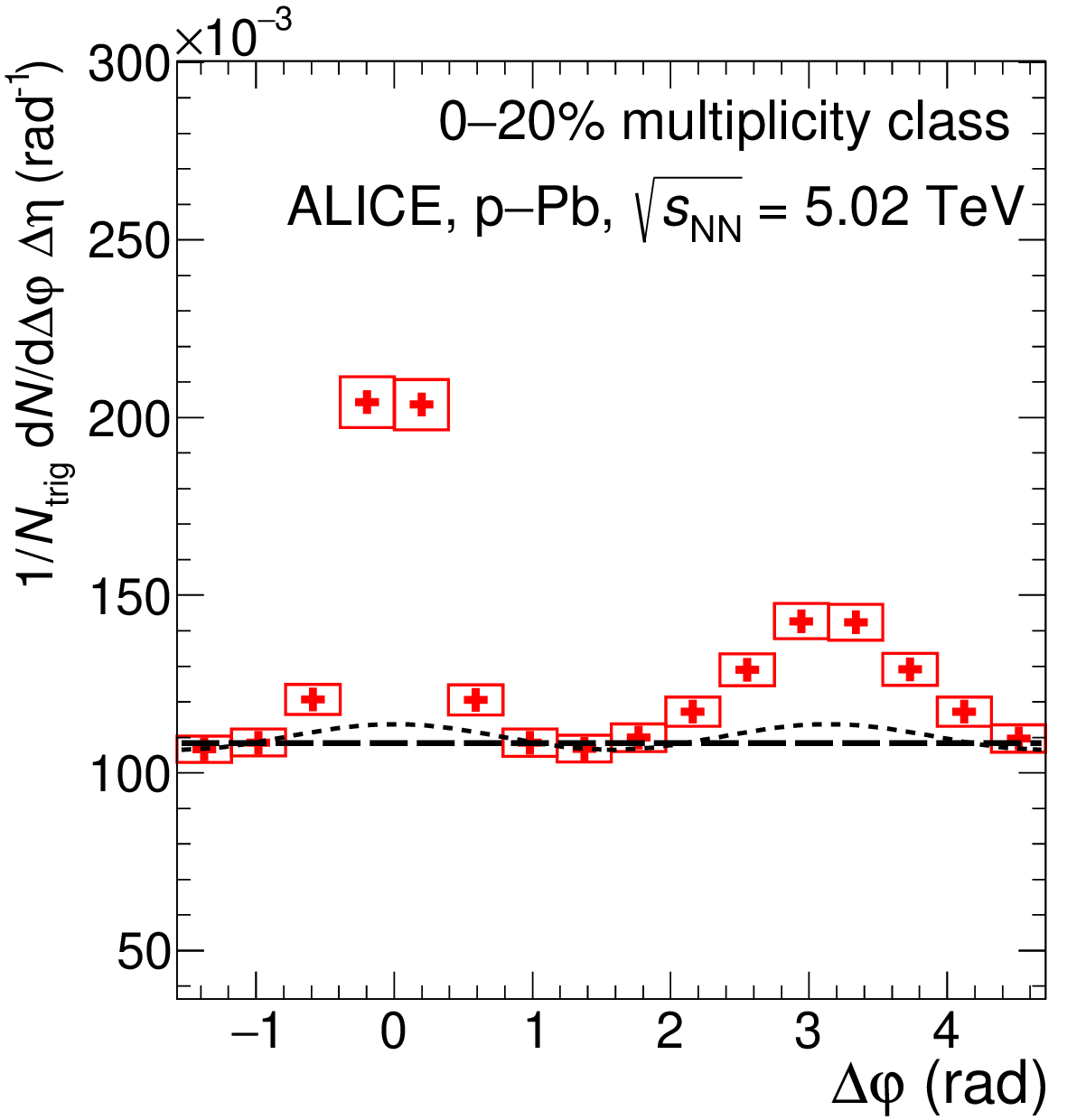}
\caption{Per-trigger dihadron $\Delta\varphi$ angular correlations integrated over the range $|\Delta\eta| < 1.2$ for the lower \pt{} range of associated hadrons, $1.5 < \pt{} < $~\SI{2.5}{\giga\eVperc} (top), and the higher \pt{} range of associated hadrons, $2.5 < \pt{} < $~\SI{4.0}{\giga\eVperc} (bottom). The three multiplicity classes are plotted in increasing order from left to right. All plots for a given associated particle \pt{} have the same axis scale, with offsets set to their respective underlying background estimation. Statistical uncertainties are represented as error bars, while systematic uncertainties are marked as open rectangles.}
\label{dphi_hh}
\end{figure}

The measured per-trigger h--$\phi$ correlation functions for the lower ($1.5 < p_{\rm T} < $~\SI{2.5}{\giga\eVperc}) and higher ($2.5 < p_{\rm T} < $~\SI{4.0}{\giga\eVperc}) associated \pt{} intervals can be seen in Fig.~\ref{dphi_hphi}. For ease of comparison, the $y$-axes of the plots are kept at a constant scale across multiplicities, and the ranges are set such that the non-jet background is aligned across all three multiplicity classes. The width of the grey band represents the systematic uncertainty on the underlying-event background, while the dotted line displays the $v_2$ background estimate, as discussed in the previous section. In both the low and high momentum correlations, the majority of h--$\phi$ pairs are found within the underlying-event region, with this fraction increasing significantly as a function of charged-particle multiplicity. The per-trigger dihadron correlation functions can be seen in Fig.~\ref{dphi_hh}. The background and $v_2$ estimates are again shown.  Similarly to the h--$\phi$ correlations, the majority of h--h pairs are found in the underlying event.

\begin{figure}[htb!]
\centering
\includegraphics[width=0.49\linewidth]{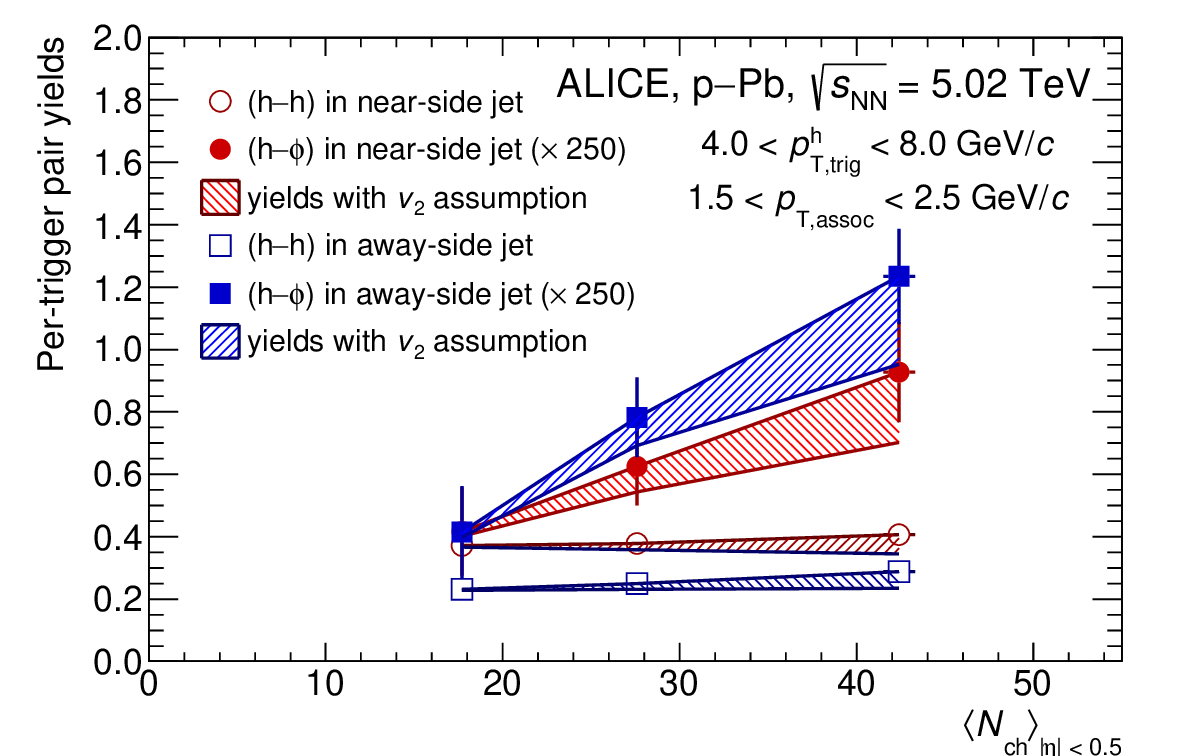}
\includegraphics[width=0.49\linewidth]{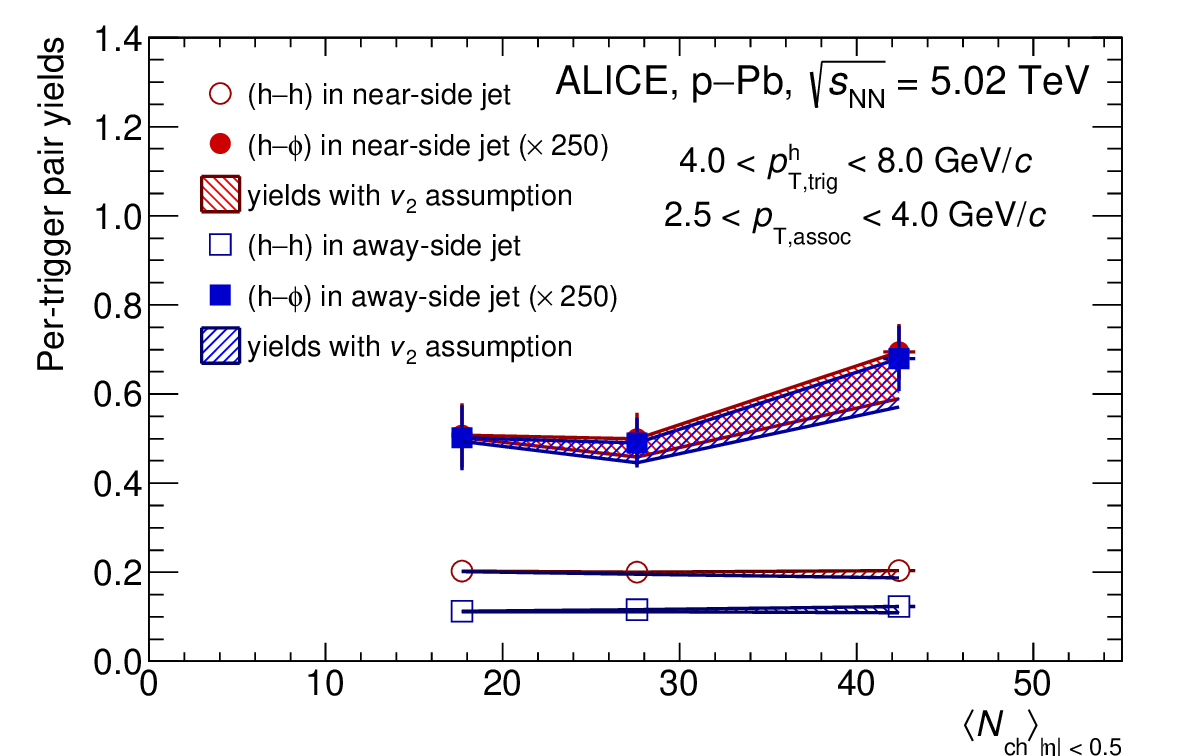}
\caption{Per-trigger pair yields within the near- and away-side jet peaks as a function of charged-particle multiplicity, for the low (left) and high (right) associated momentum bins. The shaded regions represent the deviation in the yields from using a non-zero $v_2$ when estimating the underlying-event background. Closed points are the jet yields from the h--$\phi$ correlations, while open points are the jet yields from the dihadron correlation. Only statistical uncertainties are shown. Systematic uncertainties are on the order of 10\% (15\%) of the near-side (away-side) associated $\phi$(1020) yields. As these uncertainties are correlated across multiplicity and between the near- and away-side yields, they are not plotted here.}
\label{jetyields}
\end{figure}

Yields of per-trigger correlated pairs are extracted from the correlation function for the near- and away-side jet peaks by integrating above the background in the defined $\Delta\varphi$ ranges, and are shown versus multiplicity for the lower and higher associated-particle momentum interval in Fig.~\ref{jetyields}. Here the three multiplicity percentile bins are converted to mean charged-particle multiplicity $\langle N_{\mathrm{ch}} \rangle$ in the range $|\eta|<0.5$ and $p_{\rm T} >\text{\SI{0.15}{\giga\eVperc}}$, calculated using similar techniques to Ref.~\cite{alicemult}, but with the additional requirement of a high-\pt{} trigger hadron present in the event. Due to the per-trigger scaling, these pairwise yields are good proxies for the actual $\phi$(1020) and inclusive-hadron per-trigger yields in the different azimuthal regions. The differences between the jet yields found using the non-zero $v_2$ contribution and the flat background estimates are depicted as shaded bands in the figures. While the magnitude of the $v_2$ effect on the yield grows with multiplicity, the near- and away-side yields are lowered by similar amounts, and therefore differences between the behavior of the near- and away-side yields are independent of the background $v_2$ assumption. The other systematic uncertainties, namely those associated with the tracking efficiency estimation and the $\phi$(1020) reconstruction, are correlated across multiplicity. Since all of the considered sources of systematic uncertainty are correlated between the near- and away-side yields (i.e., they have a negligible effect on the relative difference between the near- and away-side yields, and therefore no effect on the difference in trends with increasing multiplicity between the two), they are not shown in these plots.

In both associated particle momentum intervals, the dihadron jet-peak yields seen in Fig.~\ref{jetyields} show little to no change with increasing multiplicity, with the yields in the away-side jet consistently lower than for the near-side jet. However, for the h--$\phi$ jet yields, there is a distinct difference in behavior between the two momentum regions. For the higher momentum region, the near- and away-side yields are consistent within uncertainties. They also are consistent with a flat trend with respect to multiplicity, taking into account uncertainties and the $v_2$ systematic effect. In contrast, for the lower momentum region the data suggest an increase in the jet yields as a function of multiplicity. To test the increase in the low momentum jet yields, the data are fit with a flat line assuming the null hypothesis of no dependence on $\langle N_{\mathrm{ch}}\rangle$. The $\chi^2$ probability of this fit gives a p-value of 0.06 and 0.001 for the near- and away-sides, respectively. Performing the same test using the non-zero $v_2$ assumption for the UE background, which lowers the increase seen in both jet yields, gives a p-value of 0.36 and 0.03 for the near- and away-sides, respectively. Figure~\ref{jetyields} also shows a hint that at low momentum the away-side yields grow faster as a function of multiplicity than the near-side yields. Differences in near-side and away-side jet yields for high-multiplicity collisions could point to several physical effects, including differences in $\phi$(1020) production within quark-like and gluon-like jets~\cite{Ackerstaffetal.1999,stargluonjets}, as well as possible jet-medium interactions affecting strangeness production. However, more precise measurements on future larger data samples are needed to demonstrate significance in the different trends for the near- and away-side yields. 

\begin{figure}[htb!]
\centering
\subfigure[][]{
\includegraphics[width=0.8\columnwidth]{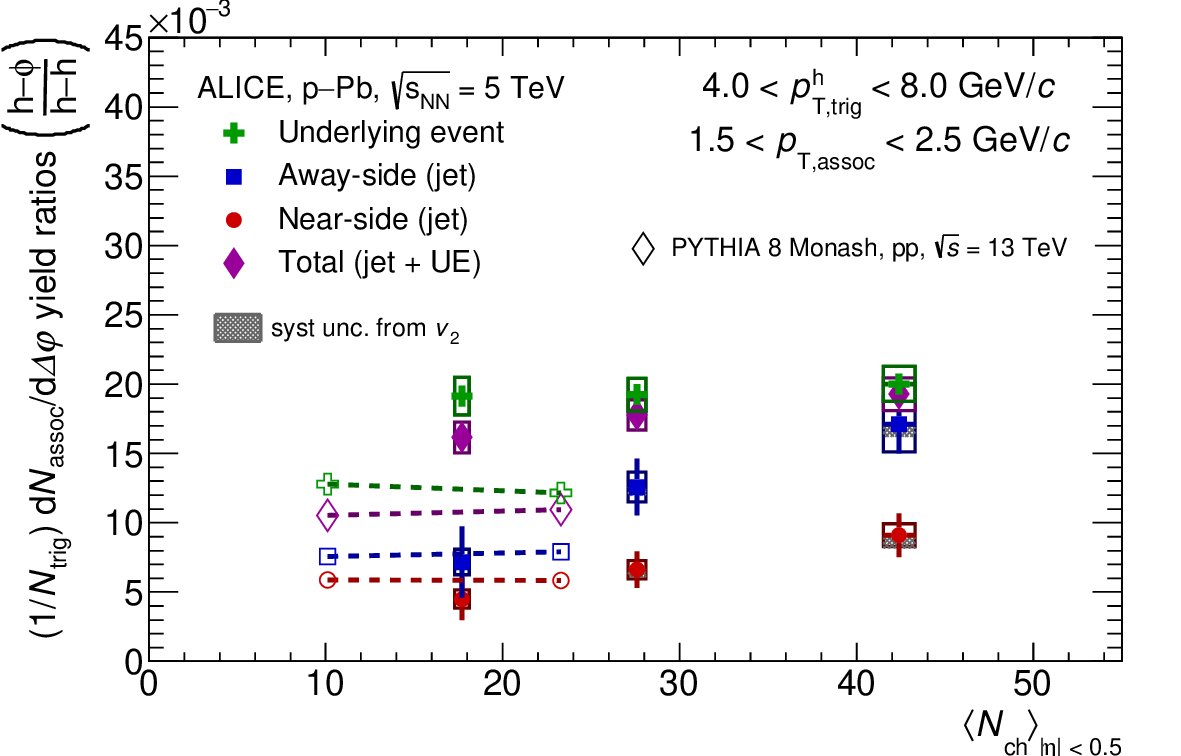}
\label{lowratiofit}
}
\subfigure[][]{
\includegraphics[width=0.8\columnwidth]{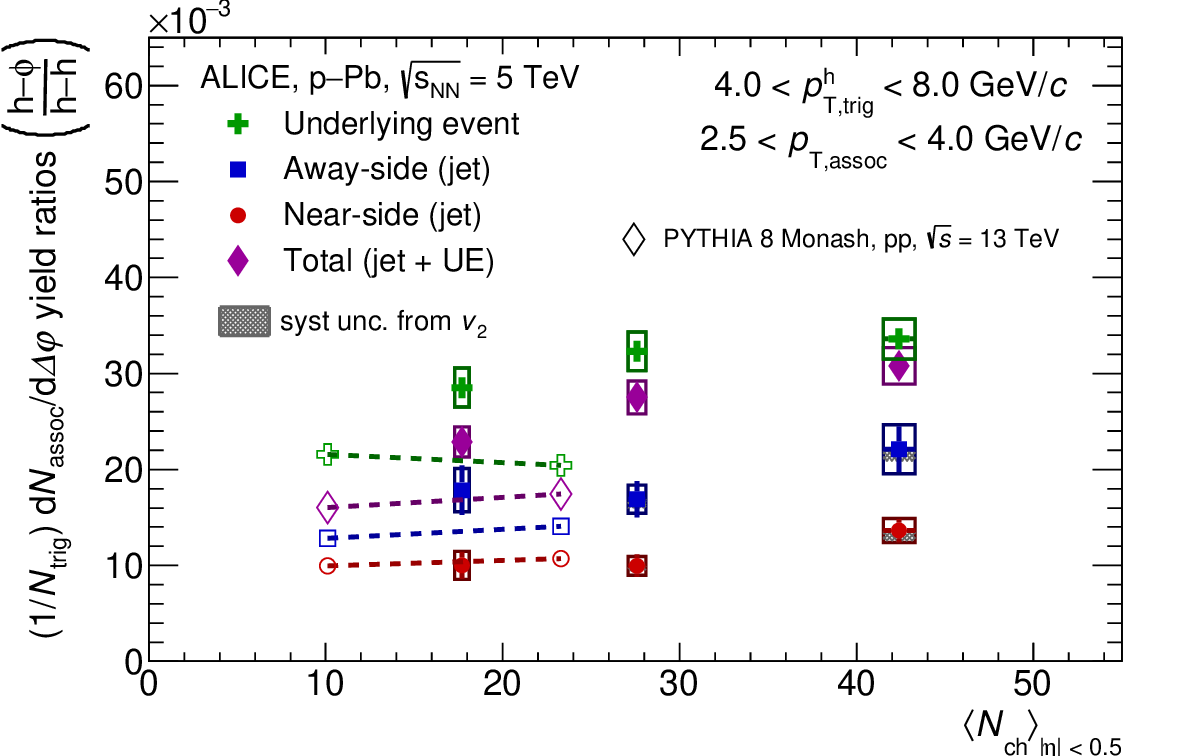}

\label{highratiofit}
}
\caption{Pairwise (h--$\phi$)/(h--h) ratio vs $\langle N_{\mathrm{ch}} \rangle$ for the lower associated momentum range, $1.5 < p_{\rm T} < $\SI{2.5}{\giga\eVperc} (top), and higher associated momentum range, $2.5 < p_{\rm T} < $\SI{4.0}{\giga\eVperc} (bottom). Statistical uncertainties are presented as error bars, while systematic uncertainties, including uncertainties on the calculated $\langle N_{\mathrm{ch}} \rangle$ value, are shown as open boxes. The additional uncertainty in the ratio due to a non-zero $v_2$ in the underlying-event background estimation is depicted as a gray band. Systematic uncertainties for the (h--h) and (h--$\phi$) yields are treated as uncorrelated when calculating the uncertainty on the ratio. Ratios from p--Pb data are depicted as solid points, while ratios from PYTHIA 8~\cite{monash} simulations of pp events are depicted as open points.}
\end{figure}

To directly compare the behavior of the $\phi$(1020) production in jets and the underlying event, it is useful to look at the pairwise yield ratio (h--$\phi$)/(h--h) as a function of multiplicity as a proxy for the $\phi/\mathrm{h}$ ratio. These ratios are plotted for the low-momentum and high-momentum regions in Fig.~\ref{lowratiofit} and Fig.~\ref{highratiofit}, respectively. The ratios for the near-side (red), away-side (blue), and the underlying event (green) show the (h--$\phi$)/(h--h) yield ratio within their respective region of the measure two-particle correlations. The total ratio (violet) represents the full (h--$\phi$)/(h--h) yield ratio. The variation between the ratios using a flat background estimation and using the non-zero $v_2$ assumption is depicted as a shaded area below the central points. This non-zero $v_2$ assumption has a small effect on the near- and away-side ratios that increases slightly with multiplicity. The effect of the $v_2$ on the underlying event ratios is negligible for all multiplicities.

From the pairwise (h--$\phi$)/(h--h) ratios, the ratio in the underlying event is observed to be significantly higher than the ratio within the jets for both momentum intervals considered here by $\approx5\sigma$, and is consistent with being multiplicity independent. The away-side ratio is higher than the near-side ratio for all multiplicities in both momentum ranges. As these pairwise ratios are approximately equal to the per-trigger $\phi/\mathrm{h}$ ratio, this suggests that $\phi$(1020) mesons have a higher probability to be produced outside of the jet compared to unidentified hadrons. It can also be seen that the total (h--$\phi$)/(h--h) ratio (violet) is predominantly determined by the underlying event production, as the majority of h--h and h--$\phi$ pairs do not come from the jets. In fact, the increase in the total (h--$\phi$)/(h--h) ratio is a consequence of the change in the contribution from jets as a function of multiplicity: for the low (high) momentum range, the fraction of total hadrons coming from jets decreases from approximately 20\% (35\%) to 10\% (20\%) as the multiplicity increases. 

The general ordering of the ratios (e.g., jet-yield ratios are consistently lower than the underlying-event yield ratios) is seen for both momentum ranges. In the lower momentum interval, the away-side ratio shows a hint of an increase of ($140 \pm 90$)\% from the lowest to the highest multiplicity, nearly reaching the values of the underlying event ratio at high multiplicity. The convergence of the ratio in the away-side jet and the underlying event at high multiplicity is suggestive of a change in the production measured within the away-side jet, from a more jet-like $\phi/\mathrm{h}$ production towards a more medium-like $\phi/\mathrm{h}$ production. The near-side jet also shows hints of an increase by ($100 \pm 75$)\%, though a larger data sample is needed to better constrain this possible increase in jet production of $\phi(1020)$. In the higher momentum interval both the near- and away-side ratios are consistent with a flat trend with increasing multiplicity. 

These results are also compared to (h--$\phi$)/(h--h) yield ratio calculated for a pp baseline using the Monash 2013 tune of the PYTHIA v8.210 Monte Carlo generator~\cite{monash}. This is used to establish the modeled behavior of strangeness production in jets and the underlying event within small collisions without an explicit deconfined phase. Similarly to data, PYTHIA 8 events are separated into two multiplicity classes, 0--20\% and 20--80\%, based on the particle production in the forward V0A acceptance. The mean multiplicity in the central pseudorapidity region is then measured for each of the two PYTHIA 8 multiplicity classes to compare directly with data. The PYTHIA 8 ratios within the near- and away-side jet are consistent with the low multiplicity p--Pb data. For both momentum bins, PYTHIA 8 under-predicts the ratio in the underlying event, as well as the total ratio, by roughly a factor of 3. Unlike the measured ratios, the ratios predicted by PYTHIA show no changes with respect to multiplicity for both the jet and underlying-event contributions.

\section{Conclusion}
\label{sec:con}
This paper presents new measurements of h--$\phi$ angular correlations in p--Pb collisions at $\sNN{} = \SI{5.02}{\tera\eV}$. The correlations are measured for associated $\phi$(1020) mesons in two transverse momentum intervals and in three multiplicity ranges in order to obtain the per-trigger yields of $\phi$(1020) mesons in the near-side peak, away-side peak, and underlying event separately. The measured yields show a distinct difference in behavior between the two momentum ranges. While the $\phi$(1020) yields measured in jets in the lower momentum interval tends to increase with increasing multiplicity, the yield measured in the higher momentum interval is roughly independent of event multiplicity. Further, the $\phi$(1020) yield measured in the lower momentum interval in the away-side jet tends to exhibit faster growth with multiplicity when compared to the near-side jet. This suggests changes to the strangeness production within the away-side jet with increasing multiplicity, though further study is needed to differentiate between possible physical causes.

Also presented are the first measurements of differentiated pairwise (h--$\phi$)/(h--h) yield ratios corresponding to the $\phi/\mathrm{h}$ ratio within jets and within the underlying event in p--Pb events. These ratios show a clear $\phi/\mathrm{h}$ ordering for both momentum ranges. The ratio is lowest in the near-side jet, followed by the away-side, and the underlying event is the highest. Interestingly, the (h--$\phi$)/(h--h) ratio within the underlying event is significantly higher than the ratio in jets even within the lowest multiplicity events, showing a $\approx 5\sigma$ difference between the underlying-event ratio and the jet ratio for the lower momentum yields. This difference in production at mid-\pt{} even in low multiplicity p--Pb events invites further study of strangeness production at lower event multiplicity. 

It can be seen that the ratios obtained in the underlying event, as well as in the jets within the higher momentum interval, do not exhibit significant dependence on multiplicity. However, the away-side jet ratio in the lower momentum range shows a slight enhancement as a function of charged-particle multiplicity. The yield ratios in jets are consistent between PYTHIA generated pp events and low multiplicity p--Pb events for both momentum ranges considered. However, the total and underlying event ratio within p--Pb data is significantly higher than the PYTHIA 8 pp baseline, across all multiplicities. This points to the enhanced $\phi$/h production within the underlying-event even in low multiplicity p--Pb events. The trends in these ratios point to the need for further studies in small systems such as high-multiplicity pp collisions, and demonstrate the usefulness of differentiated particle yields in constraining the different sources of strangeness production and enhancement.

%%%%%%%%%%%%%%%%%%%%%%%%%%%%%%%%
% end main text 
%%%%%%%%%%%%%%%%%%%%%%%%%%%%%%%%

%%%%% acknowledgements
\newenvironment{acknowledgement}{\relax}{\relax}
\begin{acknowledgement}
\section*{Acknowledgements}
% Version: 2024-05-01

The ALICE Collaboration would like to thank all its engineers and technicians for their invaluable contributions to the construction of the experiment and the CERN accelerator teams for the outstanding performance of the LHC complex.
The ALICE Collaboration gratefully acknowledges the resources and support provided by all Grid centres and the Worldwide LHC Computing Grid (WLCG) collaboration.
The ALICE Collaboration acknowledges the following funding agencies for their support in building and running the ALICE detector:
A. I. Alikhanyan National Science Laboratory (Yerevan Physics Institute) Foundation (ANSL), State Committee of Science and World Federation of Scientists (WFS), Armenia;
Austrian Academy of Sciences, Austrian Science Fund (FWF): [M 2467-N36] and Nationalstiftung f\"{u}r Forschung, Technologie und Entwicklung, Austria;
Ministry of Communications and High Technologies, National Nuclear Research Center, Azerbaijan;
Conselho Nacional de Desenvolvimento Cient\'{\i}fico e Tecnol\'{o}gico (CNPq), Financiadora de Estudos e Projetos (Finep), Funda\c{c}\~{a}o de Amparo \`{a} Pesquisa do Estado de S\~{a}o Paulo (FAPESP) and Universidade Federal do Rio Grande do Sul (UFRGS), Brazil;
Bulgarian Ministry of Education and Science, within the National Roadmap for Research Infrastructures 2020-2027 (object CERN), Bulgaria;
Ministry of Education of China (MOEC) , Ministry of Science \& Technology of China (MSTC) and National Natural Science Foundation of China (NSFC), China;
Ministry of Science and Education and Croatian Science Foundation, Croatia;
Centro de Aplicaciones Tecnol\'{o}gicas y Desarrollo Nuclear (CEADEN), Cubaenerg\'{\i}a, Cuba;
Ministry of Education, Youth and Sports of the Czech Republic, Czech Republic;
The Danish Council for Independent Research | Natural Sciences, the VILLUM FONDEN and Danish National Research Foundation (DNRF), Denmark;
Helsinki Institute of Physics (HIP), Finland;
Commissariat \`{a} l'Energie Atomique (CEA) and Institut National de Physique Nucl\'{e}aire et de Physique des Particules (IN2P3) and Centre National de la Recherche Scientifique (CNRS), France;
Bundesministerium f\"{u}r Bildung und Forschung (BMBF) and GSI Helmholtzzentrum f\"{u}r Schwerionenforschung GmbH, Germany;
General Secretariat for Research and Technology, Ministry of Education, Research and Religions, Greece;
National Research, Development and Innovation Office, Hungary;
Department of Atomic Energy Government of India (DAE), Department of Science and Technology, Government of India (DST), University Grants Commission, Government of India (UGC) and Council of Scientific and Industrial Research (CSIR), India;
National Research and Innovation Agency - BRIN, Indonesia;
Istituto Nazionale di Fisica Nucleare (INFN), Italy;
Japanese Ministry of Education, Culture, Sports, Science and Technology (MEXT) and Japan Society for the Promotion of Science (JSPS) KAKENHI, Japan;
Consejo Nacional de Ciencia (CONACYT) y Tecnolog\'{i}a, through Fondo de Cooperaci\'{o}n Internacional en Ciencia y Tecnolog\'{i}a (FONCICYT) and Direcci\'{o}n General de Asuntos del Personal Academico (DGAPA), Mexico;
Nederlandse Organisatie voor Wetenschappelijk Onderzoek (NWO), Netherlands;
The Research Council of Norway, Norway;
Pontificia Universidad Cat\'{o}lica del Per\'{u}, Peru;
Ministry of Science and Higher Education, National Science Centre and WUT ID-UB, Poland;
Korea Institute of Science and Technology Information and National Research Foundation of Korea (NRF), Republic of Korea;
Ministry of Education and Scientific Research, Institute of Atomic Physics, Ministry of Research and Innovation and Institute of Atomic Physics and Universitatea Nationala de Stiinta si Tehnologie Politehnica Bucuresti, Romania;
Ministry of Education, Science, Research and Sport of the Slovak Republic, Slovakia;
National Research Foundation of South Africa, South Africa;
Swedish Research Council (VR) and Knut \& Alice Wallenberg Foundation (KAW), Sweden;
European Organization for Nuclear Research, Switzerland;
Suranaree University of Technology (SUT), National Science and Technology Development Agency (NSTDA) and National Science, Research and Innovation Fund (NSRF via PMU-B B05F650021), Thailand;
Turkish Energy, Nuclear and Mineral Research Agency (TENMAK), Turkey;
National Academy of  Sciences of Ukraine, Ukraine;
Science and Technology Facilities Council (STFC), United Kingdom;
National Science Foundation of the United States of America (NSF) and United States Department of Energy, Office of Nuclear Physics (DOE NP), United States of America.
In addition, individual groups or members have received support from:
Czech Science Foundation (grant no. 23-07499S), Czech Republic;
European Research Council (grant no. 950692), European Union;
ICSC - Centro Nazionale di Ricerca in High Performance Computing, Big Data and Quantum Computing, European Union - NextGenerationEU;
Academy of Finland (Center of Excellence in Quark Matter) (grant nos. 346327, 346328), Finland.
    %%%%%%% done by webmaster team
\end{acknowledgement}

%%%%%%%% Bibliography 
\bibliographystyle{utphys}
\bibliography{references.bib}

\newpage
\appendix

\section{The ALICE Collaboration}
\label{app:collab}
% ALICE Collaboration author list for 2024-05-01
\begin{flushleft} 
\small

S.~Acharya\,\orcidlink{0000-0002-9213-5329}\,$^{\rm 127}$, 
D.~Adamov\'{a}\,\orcidlink{0000-0002-0504-7428}\,$^{\rm 86}$, 
A.~Agarwal$^{\rm 135}$, 
G.~Aglieri Rinella\,\orcidlink{0000-0002-9611-3696}\,$^{\rm 32}$, 
L.~Aglietta\,\orcidlink{0009-0003-0763-6802}\,$^{\rm 24}$, 
M.~Agnello\,\orcidlink{0000-0002-0760-5075}\,$^{\rm 29}$, 
N.~Agrawal\,\orcidlink{0000-0003-0348-9836}\,$^{\rm 25}$, 
Z.~Ahammed\,\orcidlink{0000-0001-5241-7412}\,$^{\rm 135}$, 
S.~Ahmad\,\orcidlink{0000-0003-0497-5705}\,$^{\rm 15}$, 
S.U.~Ahn\,\orcidlink{0000-0001-8847-489X}\,$^{\rm 71}$
% ...

\section*{Affiliation Notes}

$^{\rm I}$ Deceased\\
$^{\rm II}$ Also at: Max-Planck-Institut fur Physik, Munich, Germany\\
$^{\rm III}$ Also at: Italian National Agency for New Technologies, Energy and Sustainable Economic Development (ENEA), Bologna, Italy\\
$^{\rm IV}$ Also at: Dipartimento DET del Politecnico di Torino, Turin, Italy\\
$^{\rm V}$ Also at: Yildiz Technical University, Istanbul, T\"{u}rkiye\\
$^{\rm VI}$ Also at: Department of Applied Physics, Aligarh Muslim University, Aligarh, India\\
$^{\rm VII}$ Also at: Institute of Theoretical Physics, University of Wroclaw, Poland\\
$^{\rm VIII}$ Also at: An institution covered by a cooperation agreement with CERN\\

\section*{Collaboration Institutes}

$^{1}$ A.I. Alikhanyan National Science Laboratory (Yerevan Physics Institute) Foundation, Yerevan, Armenia\\
$^{2}$ AGH University of Krakow, Cracow, Poland\\
$^{3}$ Bogolyubov Institute for Theoretical Physics, National Academy of Sciences of Ukraine, Kiev, Ukraine\\
$^{4}$ Bose Institute, Department of Physics  and Centre for Astroparticle Physics and Space Science (CAPSS), Kolkata, India\\
$^{5}$ California Polytechnic State University, San Luis Obispo, California, United States\\
$^{6}$ Central China Normal University, Wuhan, China\\
$^{7}$ Centro de Aplicaciones Tecnol\'{o}gicas y Desarrollo Nuclear (CEADEN), Havana, Cuba\\
$^{8}$ Centro de Investigaci\'{o}n y de Estudios Avanzados (CINVESTAV), Mexico City and M\'{e}rida, Mexico\\
$^{9}$ Chicago State University, Chicago, Illinois, United States\\
$^{10}$ China Institute of Atomic Energy, Beijing, China\\
$^{11}$ China University of Geosciences, Wuhan, China\\
$^{12}$ Chungbuk National University, Cheongju, Republic of Korea\\
$^{13}$ Comenius University Bratislava, Faculty of Mathematics, Physics and Informatics, Bratislava, Slovak Republic\\
$^{14}$ Creighton University, Omaha, Nebraska, United States\\
$^{15}$ Department of Physics, Aligarh Muslim University, Aligarh, India\\
$^{16}$ Department of Physics, Pusan National University, Pusan, Republic of Korea\\
$^{17}$ Department of Physics, Sejong University, Seoul, Republic of Korea\\
$^{18}$ Department of Physics, University of California, Berkeley, California, United States\\
$^{19}$ Department of Physics, University of Oslo, Oslo, Norway\\
$^{20}$ Department of Physics and Technology, University of Bergen, Bergen, Norway\\
$^{21}$ Dipartimento di Fisica, Universit\`{a} di Pavia, Pavia, Italy\\
$^{22}$ Dipartimento di Fisica dell'Universit\`{a} and Sezione INFN, Cagliari, Italy\\
$^{23}$ Dipartimento di Fisica dell'Universit\`{a} and Sezione INFN, Trieste, Italy\\
$^{24}$ Dipartimento di Fisica dell'Universit\`{a} and Sezione INFN, Turin, Italy\\
$^{25}$ Dipartimento di Fisica e Astronomia dell'Universit\`{a} and Sezione INFN, Bologna, Italy\\
$^{26}$ Dipartimento di Fisica e Astronomia dell'Universit\`{a} and Sezione INFN, Catania, Italy\\
$^{27}$ Dipartimento di Fisica e Astronomia dell'Universit\`{a} and Sezione INFN, Padova, Italy\\
$^{28}$ Dipartimento di Fisica `E.R.~Caianiello' dell'Universit\`{a} and Gruppo Collegato INFN, Salerno, Italy\\
$^{29}$ Dipartimento DISAT del Politecnico and Sezione INFN, Turin, Italy\\
$^{30}$ Dipartimento di Scienze MIFT, Universit\`{a} di Messina, Messina, Italy\\
$^{31}$ Dipartimento Interateneo di Fisica `M.~Merlin' and Sezione INFN, Bari, Italy\\
$^{32}$ European Organization for Nuclear Research (CERN), Geneva, Switzerland\\
$^{33}$ Faculty of Electrical Engineering, Mechanical Engineering and Naval Architecture, University of Split, Split, Croatia\\
$^{34}$ Faculty of Engineering and Science, Western Norway University of Applied Sciences, Bergen, Norway\\
$^{35}$ Faculty of Nuclear Sciences and Physical Engineering, Czech Technical University in Prague, Prague, Czech Republic\\
$^{36}$ Faculty of Physics, Sofia University, Sofia, Bulgaria\\
$^{37}$ Faculty of Science, P.J.~\v{S}af\'{a}rik University, Ko\v{s}ice, Slovak Republic\\
$^{38}$ Frankfurt Institute for Advanced Studies, Johann Wolfgang Goethe-Universit\"{a}t Frankfurt, Frankfurt, Germany\\
$^{39}$ Fudan University, Shanghai, China\\
$^{40}$ Gangneung-Wonju National University, Gangneung, Republic of Korea\\
$^{41}$ Gauhati University, Department of Physics, Guwahati, India\\
$^{42}$ Helmholtz-Institut f\"{u}r Strahlen- und Kernphysik, Rheinische Friedrich-Wilhelms-Universit\"{a}t Bonn, Bonn, Germany\\
$^{43}$ Helsinki Institute of Physics (HIP), Helsinki, Finland\\
$^{44}$ High Energy Physics Group,  Universidad Aut\'{o}noma de Puebla, Puebla, Mexico\\
$^{45}$ Horia Hulubei National Institute of Physics and Nuclear Engineering, Bucharest, Romania\\
$^{46}$ HUN-REN Wigner Research Centre for Physics, Budapest, Hungary\\
$^{47}$ Indian Institute of Technology Bombay (IIT), Mumbai, India\\
$^{48}$ Indian Institute of Technology Indore, Indore, India\\
$^{49}$ INFN, Laboratori Nazionali di Frascati, Frascati, Italy\\
$^{50}$ INFN, Sezione di Bari, Bari, Italy\\
$^{51}$ INFN, Sezione di Bologna, Bologna, Italy\\
$^{52}$ INFN, Sezione di Cagliari, Cagliari, Italy\\
$^{53}$ INFN, Sezione di Catania, Catania, Italy\\
$^{54}$ INFN, Sezione di Padova, Padova, Italy\\
$^{55}$ INFN, Sezione di Pavia, Pavia, Italy\\
$^{56}$ INFN, Sezione di Torino, Turin, Italy\\
$^{57}$ INFN, Sezione di Trieste, Trieste, Italy\\
$^{58}$ Inha University, Incheon, Republic of Korea\\
$^{59}$ Institute for Gravitational and Subatomic Physics (GRASP), Utrecht University/Nikhef, Utrecht, Netherlands\\
$^{60}$ Institute of Experimental Physics, Slovak Academy of Sciences, Ko\v{s}ice, Slovak Republic\\
$^{61}$ Institute of Physics, Homi Bhabha National Institute, Bhubaneswar, India\\
$^{62}$ Institute of Physics of the Czech Academy of Sciences, Prague, Czech Republic\\
$^{63}$ Institute of Space Science (ISS), Bucharest, Romania\\
$^{64}$ Institut f\"{u}r Kernphysik, Johann Wolfgang Goethe-Universit\"{a}t Frankfurt, Frankfurt, Germany\\
$^{65}$ Instituto de Ciencias Nucleares, Universidad Nacional Aut\'{o}noma de M\'{e}xico, Mexico City, Mexico\\
$^{66}$ Instituto de F\'{i}sica, Universidade Federal do Rio Grande do Sul (UFRGS), Porto Alegre, Brazil\\
$^{67}$ Instituto de F\'{\i}sica, Universidad Nacional Aut\'{o}noma de M\'{e}xico, Mexico City, Mexico\\
$^{68}$ iThemba LABS, National Research Foundation, Somerset West, South Africa\\
$^{69}$ Jeonbuk National University, Jeonju, Republic of Korea\\
$^{70}$ Johann-Wolfgang-Goethe Universit\"{a}t Frankfurt Institut f\"{u}r Informatik, Fachbereich Informatik und Mathematik, Frankfurt, Germany\\
$^{71}$ Korea Institute of Science and Technology Information, Daejeon, Republic of Korea\\
$^{72}$ KTO Karatay University, Konya, Turkey\\
$^{73}$ Laboratoire de Physique Subatomique et de Cosmologie, Universit\'{e} Grenoble-Alpes, CNRS-IN2P3, Grenoble, France\\
$^{74}$ Lawrence Berkeley National Laboratory, Berkeley, California, United States\\
$^{75}$ Lund University Department of Physics, Division of Particle Physics, Lund, Sweden\\
$^{76}$ Nagasaki Institute of Applied Science, Nagasaki, Japan\\
$^{77}$ Nara Women{'}s University (NWU), Nara, Japan\\
$^{78}$ National and Kapodistrian University of Athens, School of Science, Department of Physics , Athens, Greece\\
$^{79}$ National Centre for Nuclear Research, Warsaw, Poland\\
$^{80}$ National Institute of Science Education and Research, Homi Bhabha National Institute, Jatni, India\\
$^{81}$ National Nuclear Research Center, Baku, Azerbaijan\\
$^{82}$ National Research and Innovation Agency - BRIN, Jakarta, Indonesia\\
$^{83}$ Niels Bohr Institute, University of Copenhagen, Copenhagen, Denmark\\
$^{84}$ Nikhef, National institute for subatomic physics, Amsterdam, Netherlands\\
$^{85}$ Nuclear Physics Group, STFC Daresbury Laboratory, Daresbury, United Kingdom\\
$^{86}$ Nuclear Physics Institute of the Czech Academy of Sciences, Husinec-\v{R}e\v{z}, Czech Republic\\
$^{87}$ Oak Ridge National Laboratory, Oak Ridge, Tennessee, United States\\
$^{88}$ Ohio State University, Columbus, Ohio, United States\\
$^{89}$ Physics department, Faculty of science, University of Zagreb, Zagreb, Croatia\\
$^{90}$ Physics Department, Panjab University, Chandigarh, India\\
$^{91}$ Physics Department, University of Jammu, Jammu, India\\
$^{92}$ Physics Program and International Institute for Sustainability with Knotted Chiral Meta Matter (SKCM2), Hiroshima University, Hiroshima, Japan\\
$^{93}$ Physikalisches Institut, Eberhard-Karls-Universit\"{a}t T\"{u}bingen, T\"{u}bingen, Germany\\
$^{94}$ Physikalisches Institut, Ruprecht-Karls-Universit\"{a}t Heidelberg, Heidelberg, Germany\\
$^{95}$ Physik Department, Technische Universit\"{a}t M\"{u}nchen, Munich, Germany\\
$^{96}$ Politecnico di Bari and Sezione INFN, Bari, Italy\\
$^{97}$ Research Division and ExtreMe Matter Institute EMMI, GSI Helmholtzzentrum f\"ur Schwerionenforschung GmbH, Darmstadt, Germany\\
$^{98}$ Saga University, Saga, Japan\\
$^{99}$ Saha Institute of Nuclear Physics, Homi Bhabha National Institute, Kolkata, India\\
$^{100}$ School of Physics and Astronomy, University of Birmingham, Birmingham, United Kingdom\\
$^{101}$ Secci\'{o}n F\'{\i}sica, Departamento de Ciencias, Pontificia Universidad Cat\'{o}lica del Per\'{u}, Lima, Peru\\
$^{102}$ Stefan Meyer Institut f\"{u}r Subatomare Physik (SMI), Vienna, Austria\\
$^{103}$ SUBATECH, IMT Atlantique, Nantes Universit\'{e}, CNRS-IN2P3, Nantes, France\\
$^{104}$ Sungkyunkwan University, Suwon City, Republic of Korea\\
$^{105}$ Suranaree University of Technology, Nakhon Ratchasima, Thailand\\
$^{106}$ Technical University of Ko\v{s}ice, Ko\v{s}ice, Slovak Republic\\
$^{107}$ The Henryk Niewodniczanski Institute of Nuclear Physics, Polish Academy of Sciences, Cracow, Poland\\
$^{108}$ The University of Texas at Austin, Austin, Texas, United States\\
$^{109}$ Universidad Aut\'{o}noma de Sinaloa, Culiac\'{a}n, Mexico\\
$^{110}$ Universidade de S\~{a}o Paulo (USP), S\~{a}o Paulo, Brazil\\
$^{111}$ Universidade Estadual de Campinas (UNICAMP), Campinas, Brazil\\
$^{112}$ Universidade Federal do ABC, Santo Andre, Brazil\\
$^{113}$ Universitatea Nationala de Stiinta si Tehnologie Politehnica Bucuresti, Bucharest, Romania\\
$^{114}$ University of Cape Town, Cape Town, South Africa\\
$^{115}$ University of Derby, Derby, United Kingdom\\
$^{116}$ University of Houston, Houston, Texas, United States\\
$^{117}$ University of Jyv\"{a}skyl\"{a}, Jyv\"{a}skyl\"{a}, Finland\\
$^{118}$ University of Kansas, Lawrence, Kansas, United States\\
$^{119}$ University of Liverpool, Liverpool, United Kingdom\\
$^{120}$ University of Science and Technology of China, Hefei, China\\
$^{121}$ University of South-Eastern Norway, Kongsberg, Norway\\
$^{122}$ University of Tennessee, Knoxville, Tennessee, United States\\
$^{123}$ University of the Witwatersrand, Johannesburg, South Africa\\
$^{124}$ University of Tokyo, Tokyo, Japan\\
$^{125}$ University of Tsukuba, Tsukuba, Japan\\
$^{126}$ Universit\"{a}t M\"{u}nster, Institut f\"{u}r Kernphysik, M\"{u}nster, Germany\\
$^{127}$ Universit\'{e} Clermont Auvergne, CNRS/IN2P3, LPC, Clermont-Ferrand, France\\
$^{128}$ Universit\'{e} de Lyon, CNRS/IN2P3, Institut de Physique des 2 Infinis de Lyon, Lyon, France\\
$^{129}$ Universit\'{e} de Strasbourg, CNRS, IPHC UMR 7178, F-67000 Strasbourg, France, Strasbourg, France\\
$^{130}$ Universit\'{e} Paris-Saclay, Centre d'Etudes de Saclay (CEA), IRFU, D\'{e}partment de Physique Nucl\'{e}aire (DPhN), Saclay, France\\
$^{131}$ Universit\'{e}  Paris-Saclay, CNRS/IN2P3, IJCLab, Orsay, France\\
$^{132}$ Universit\`{a} degli Studi di Foggia, Foggia, Italy\\
$^{133}$ Universit\`{a} del Piemonte Orientale, Vercelli, Italy\\
$^{134}$ Universit\`{a} di Brescia, Brescia, Italy\\
$^{135}$ Variable Energy Cyclotron Centre, Homi Bhabha National Institute, Kolkata, India\\
$^{136}$ Warsaw University of Technology, Warsaw, Poland\\
$^{137}$ Wayne State University, Detroit, Michigan, United States\\
$^{138}$ Yale University, New Haven, Connecticut, United States\\
$^{139}$ Yonsei University, Seoul, Republic of Korea\\
$^{140}$  Zentrum  f\"{u}r Technologie und Transfer (ZTT), Worms, Germany\\
$^{141}$ Affiliated with an institute covered by a cooperation agreement with CERN\\
$^{142}$ Affiliated with an international laboratory covered by a cooperation agreement with CERN.\\

\end{flushleft} 
  %%%%%%% done by webmaster team
\end{document}